\documentclass[a4paper,11pt,graphicx]{article}

\usepackage[english]{babel}
\usepackage{amssymb}
\usepackage{amsmath}
\usepackage[dvips]{graphics}
\usepackage[dvips]{graphicx}

\newtheorem{theorem}{{\bf Theorem}}

\newtheorem{lemma}{Lemma}

\newenvironment{proof}[1][Proof]{\textbf{#1.} }{\ \rule{0.5em}{0.5em}}

\numberwithin{equation}{section}
\numberwithin{figure}{section}

\newcommand{\benumerate}{\begin{enumerate}}
\newcommand{\eenumerate}{\end{enumerate}}

\newcommand{\bitemize}{\begin{itemize}}

\newcommand{\eitemize}{\end{itemize}}
\newcommand{\der}[2]{\frac{\partial #1}{\partial #2}}

\newcommand{\dersec}[2]{\frac{\partial^{2} #1}{\partial #2^{2}}}
\newcommand{\derthree}[2]{\frac{\partial^{3} #1}{\partial #2^{3}}}

\newcommand{\dermixd}[3]{\frac{\partial^{2} #1}{\partial #2 ~\partial #3}}

\newcommand{\dertot}[2]{\frac{d #1}{d #2}}

\newcommand{\simbon}[2]{\mathop{\rm #1}\limits_{#2}}

\newcommand{\paperspie}[7]{#1, ``#2'', {\it #3}, {\bf #4}, pp. #5-#6, #7}

\newcommand{\paperbu}[7]{#1, {\it #2}, #3, {\bf #4}, (#5) #6-#7}
\newcommand{\bookbu}[5]{#1, {\it #2}, #3, #4, #5}

\begin{document}

\date{}

\title{
High frequency integrable regimes in nonlocal nonlinear optics.}

\author{Antonio Moro and Boris Konopelchenko \\
\small{Dipartimento di Fisica dell'Universit\`{a} di Lecce} \\
\small{and INFN, Sezione di Lecce, I-73100 Lecce, Italy}\\
\small{E-mail: antonio.moro@le.infn.it, konopel@le.infn.it}}

\maketitle

\begin{abstract}
We consider an integrable model which describes light beams
propagating in nonlocal nonlinear media of Cole-Cole type. The
model is derived as high frequency limit of both Maxwell equations
and the nonlocal nonlinear Schr\"odinger equation. We demonstrate
that for a general form of nonlinearity there exist selfguided
light beams. In high frequency limit nonlocal perturbations can be
seen as a class of phase deformation along one direction. We study
in detail nonlocal perturbations described by the dispersionless
Veselov-Novikov (dVN) hierarchy. The dVN hierarchy is analyzed by
the reduction method based on symmetry constraints and by the
quasiclassical $\bar{\partial}-$dressing method. Quasiclassical
$\bar{\partial}-$dressing method reveals a connection between
nonlocal nonlinear
geometric optics and the theory of quasiconformal mappings of the plane.\\

\end{abstract}

%\tableofcontents

\section{Introduction.}
\label{sec_intro}
%\begin{figure}
%\label{argand}
%\centerline{\includegraphics[width=3cm]{prova3.eps}}
%\end{figure}
The {\em optics} studies phenomena of the propagation of the
electromagnetic waves through a {\em dielectric} medium in absence
of currents and charges~\cite{Born}. In such a case the Maxwell
equations assume the form
\begin{subequations}
\label{maxwell_free}
\begin{align}
\label{maxwell1}
\nabla \wedge {\bf H} -
\der{{\bf D}}{t} &= 0 \\
\label{maxwell2}
\nabla \wedge {\bf E} + \der{{\bf B}}{t} &= 0  \\
\label{maxwell3}
\nabla \cdot {\bf D}  &= 0 \\
\label{maxwell4} \nabla \cdot {\bf B} &= 0.
\end{align}
\end{subequations}
where $x,y,z$ are the spatial coordinates, $\nabla =
\left(\partial_{x},\partial_{y},\partial_{z} \right)$ is the
gradient and $t$ is the time. For sake of simplicity we have set
the light speed $c = 1$. The vectors ${\bf E}$ and ${\bf B}$ are
the electric and magnetic fields respectively, while the
displacement vector ${\bf D}$ and the magnetic induction ${\bf H}$
contain the information about the response of the medium when an
external electromagnetic field is applied. In general, ${\bf D}$
and {\bf H} are certain functions of ${\bf E}$ and ${\bf B}$ and
they are specified by the so-called {\em constitutive relations}.
In several physically meaningful cases they can be written down as
follows
\begin{equation}
\label{constitutive} {\bf D} = \varepsilon {\bf E} \quad{} {\bf B}
= \mu {\bf H},
\end{equation}
where {\em electric permittivity} $\varepsilon$ and the {\em
magnetic permeability} $\mu$ depend on the coordinates and on the
fields. In the case in which $\varepsilon$ and $\mu$ depend at
most on the coordinates, the Maxwell equations are linear and
provides us with an exhaustive description of a very broad class
of phenomena in physics~\cite{Born,Jackson,Landau}.

The {\em nonlinear optics} deals with the class of media such that
$\varepsilon$ and $\mu$ depend on the fields. In this case, the
Maxwell equations are nonlinear and construction of their
solutions is, in general, a very challenging problem. In the
following, we will focus only on the electric part of the field,
assuming the magnetic response to be negligible, i.e. $\mu =1$.
This condition is realized in many experimental situations.

By definition, a dielectric medium is referred to as {\em
nonlinear} if the dielectric function $\varepsilon$ is a certain
function of the electric field $\varepsilon = \varepsilon
\left({\bf E} \right)$.

The solution of the Maxwell equations for an arbitrary form of the
function $\varepsilon$ is, of course, a hard problem.
Nevertheless, several very interesting cases, such as quadratic
and cubic nonlinearities, are amenable by exact methods. Moreover,
they are connected with a relevant phenomenology such as the
higher harmonic generation (quadratic nonlinearity) and the
soliton production (cubic nonlinearity)~\cite{Boyd}.

In the present paper we consider a model for light beams
propagation in the limit of high frequency. This model can be
derived both as high frequency limit of the Maxwell equations and
from the nonlinear Schr\"odinger (NLS) equation which describes
paraxial light beams. We will show that it is reasonable to assume
the nonlocality to be weak because of the quickly oscillating
fields. In particular, we can separate the pure nonlinear
contribution from the higher orders nonlocal perturbations.
Leading order is represented by the standard eikonal equation.
Note that in the standard theory higher order terms contain the
amplitude of the electric field as well as the phase $S$ and
consequently in virtue of nonlocality and nonlinearity they are
rather complicated. Nevertheless, there exist a type of media
where it is possible to consider a set of nonlocal perturbations
along one direction, say $z$, which are not mixed up with the wave
contributions. These are the Cole-Cole media. They are
characterized by the Cole-Cole dispersion law for the dielectric
function of the form $\varepsilon = \varepsilon_{0} +
O(\omega^{-\alpha})$ for $0<\alpha<1$ and $\omega \to \infty$. We
would like to stress that a large variety of solid and liquid
polar media obeys the Cole-Cole dispersion law~\cite{ColeCole}. In
this case we perform an asymptotic expansion of all fields with
respect to the small parameter $\omega^{-\alpha}$ separating first
nonlocal correction to the phase $S$ from the higher order wave
corrections. In our derivations a suitable slow dependence along
$z-$direction is assumed. Nonlocal corrections are described by
polynomials in $S_{x}$ and $S_{y}$. In particular, a one-to-one
correspondence among nonlocality and polynomials degrees is
realized.

In the $\omega \to \infty$ limit separation between the nonlinear
response and the nonlocal effects along $z$ suggest to study first
the properties of the light beam on the $xy-$plane ({\it
transverse equations}) and then the ``evolution" along $z$.
Transverse equations are the two-dimensional eikonal equation and
the geometric optics limit of the Poynting vector conservation
law. It is shown that for a general class of nonlinear responses
obeying the so-called ellipticity condition there exists
self-guided light beams. Moreover, several light beam features are
determined in terms of the properties of quasiconformal mappings
on the plane via the Beltrami equation. Propagation of helicoidal
wavefronts and its first degree nonlocal perturbations are also
discussed. Thereafter, we focus on the special class of
perturbations which preserve the phase inversion symmetry $S \to
-S$ similar to the eikonal equation. Such perturbations are shown
to be described by an infinite set of integrable nonlinear partial
differential equations (PDEs). It is the dispersionless
Veselov-Novikov (dVN) hierarchy. First non-trivial equation of the
hierarchy can be formally obtained by the slow variable expansion
of the Veselov-Novikov equation, which has been introduced
in~\cite{VNoriginal} as $(2+1)D$ integrable generalization of
Korteweg-de-Vries equation. The dVN hierarchy is of interest since
it is treatable by different approaches. More specifically, we
discuss the reduction method based on the symmetry constraints. It
allows us to reduce the dVN equation to a $1+1-$dimensional
hydrodynamic-type system and it is effective for construction of
explicit solutions. Moreover, we study the eikonal equation and
the dVN hierarchy using the quasiclassical
$\bar{\partial}-$dressing method. It provides us with a general
approach to construct and analyze the whole hierarchy.

It is worth to note that the quasiclassical
$\bar{\partial}-$dressing method establishes a remarkable
connection between the nonlocal nonlinear geometric optics and the
theory of quasiconformal mappings of the plane.

The paper is organized as follows. The model under consideration
is derived from the Maxwell equations in
Section~\ref{sec_generalmodel} and starting with NLS equation in
the Section~\ref{sec_paraxial}. In the
Section~\ref{sec_transverse} we analyze the transverse equations
in connection with the Beltrami equation and minimal surfaces
equation which gives the helicoidal wavefronts. Nonlocal
perturbations described by the dVN hierarchy are derived in the
Section~\ref{sec_dVN_perturbation}. After a short review about
integrable system in the Section~\ref{sec_integrable} we discuss
the hydrodynamic-type reductions obtained using the symmetry
constraints in Section~\ref{sec_sym_contraints} and the
application of the quasiclassical $\bar{\partial}-$dressing method
to the eikonal equation and to the dVN hierarchy in
Section~\ref{sec_QC_Dbar}. Some concluding remarks close the
paper.

\section{The general model.}
\label{sec_generalmodel} Let us consider a medium with the
magnetic permeability $\mu = 1$. In such a case the Maxwell
equations~(\ref{maxwell_free}) imply that
\begin{equation}
\label{maxwell_wave} \nabla \wedge \nabla \wedge {\bf E} +
\dersec{{\bf D}}{t} = 0
\end{equation}
For time oscillating solutions, of the form
\begin{align}
{\bf E}(x,y,z,t) &= {\bf E}(x,y,z) \; e^{i \omega t} \nonumber \\
{\bf D}(x,y,z,t) &= {\bf D}(x,y,z) \; e^{i \omega t}, \nonumber
\end{align}
equation~(\ref{maxwell_wave}) looks like
\begin{equation}
\label{maxwell_wave2} \nabla \wedge \nabla \wedge {\bf E} -
\omega^{2} {\bf D} = 0,
\end{equation}
or, equivalently,
\begin{equation}
\label{wave_general} \nabla^{2} {\bf E} + \omega^{2} {\bf D} -
\nabla \left(\nabla \cdot {\bf E} \right) = 0.
\end{equation}
Once the constitutive relation for ${\bf D}$ is assigned
equation~(\ref{wave_general}) determines the electric
 component of the field.

The constitutive relation depends on the physical properties of
the medium. Let us assume that the displacement vector ${\bf D}$
can be splitted in a local part ${\bf D}_{L}$ and a nonlocal one
${\bf D}_{N}$ as follows
\begin{equation}
\label{def_displacement} {\bf D} = {\bf D}_{L} +  {\bf D}_{N}
\end{equation}
where
\begin{align}
\label{DL} {\bf D}_{L} &= \varepsilon \left(I(\bf x) \right) \;{\bf E}(\bf x) \\
\label{DN} {\bf D}_{N} &= \int_{\mathbb{R}^{3}} R \left({\bf x}' -
{\bf x},a (\omega) \right) \; N \left(I({\bf x'}) \right)\;{\bf
E}({\bf x'}) \; d{\bf x'}.
\end{align}
The quantity $I = \left |{\bf E}_{0} \right|^{2}$ is the intensity
of the electric field, $\varepsilon (I)$ and $N \left(I \right)$
are, respectively, the {\em local} and {\em nonlocal nonlinear
responses} and the function $R({\bf x} - {\bf x'})$ is the
nonlocal distribution. The parameter $a (\omega)$ governs the
width of the nonlocal response in different regimes of the
frequency. In particular, in the high frequency limit it is
reasonable to assume that the external field is not resonant with
the proper oscillations of the particles of the medium in such a
way that the response becomes local, i.e.
\begin{equation}
\label{alpha_limit} \lim_{\omega \to \infty} R \left({\bf x}' -
{\bf x}, a (\omega) \right) = \delta \left({\bf x }' - {\bf
x}\right),
\end{equation}
where $\delta ({\bf x}' - {\bf x})$ denotes the Dirac
$\delta-$function. In order to explain this behavior, one can
consider a naive model where the single particle of the medium is
described by a one dimensional forced oscillator
\begin{equation}
\label{forced} \ddot{x} + \omega_{0} x = f_{0} \sin \left(\omega t
\right).
\end{equation}
As usual, the dot denotes the total time derivative, $\omega_{0}$
is the proper oscillation frequency of the particle. The left hand
side contains the forcing term represented by a linearly polarized
electric field oscillating with frequency $\omega$. The solution
of equation~(\ref{forced}) is of the form
\begin{gather}
\label{forced_solution} x = \left \{
\begin{aligned}
&c_{1} \sin \left(\omega_{0} t \right) + c_{2} \cos
\left(\omega_{0} t \right) + \frac{f_{0}}{\omega_{0}^{2} -
\omega^{2}} \sin\left(\omega t \right) \quad{} &\omega_{0} \neq
\omega \\
&c_{1} \sin \left(\omega_{0} t \right) + c_{2} \cos
\left(\omega_{0} t \right) - \frac{f_{0}}{2 \omega_{0}} \: t \:
\cos \left( \omega_{0} t \right) \quad{} &\omega_{0} = \omega
\end{aligned}
\right.
\end{gather}
We remind that we are interested in the non-resonant regime
$\omega_{0} \neq \omega$. In this case, it is easy to see that for
$\omega \to \infty$ the forcing contribution disappears and the
particle behaves as the harmonic oscillator. In this regime one
expects that the effect of the external field, up to higher
corrections, does not propagate far from the point considered and
the response tends to be as local as much $\omega$ is larger.

Substituting the expression~({\ref{def_displacement}}) into
equation~(\ref{maxwell3}) one gets
\begin{equation}
\label{divergenceE} \nabla \cdot {\bf E} = - \nabla \left(\log
\varepsilon \right) \cdot {\bf E} -  \frac{\nabla \cdot {\bf
D}_{N}}{\varepsilon}.
\end{equation}
Thus, equation~(\ref{wave_general}) takes the form
\begin{equation}
\label{wave_particular} \nabla^{2} {\bf E} + \omega^{2} \left(
\varepsilon {\bf E} +  {\bf D}_{N}  \right) - \nabla \left (-
\nabla \log \varepsilon \cdot {\bf E} -  \frac{\nabla \cdot {\bf
D}_{N}}{\varepsilon} \right) = 0.
\end{equation}
Let us introduce a general model of weak nonlocality. It is given
by a nonlocal distribution function of the following form
\begin{align}
\label{weak_distribution} R\left({\bf x} - {\bf x}' \right) &=
\rho_{lmn}\left({\bf x} \right) \delta^{(l,m,n)} \left({\bf
x}-{\bf x}' \right)
\end{align}
where the sum on the repeated indices is assumed. The
distributions $\delta^{(l,m,n)}$ are the Dirac $\delta-$function
derivatives. They are standardly defined via
\begin{equation}
\label{dirac_def} \int_{\mathbb{R}^{3}} \delta^{(l,m,n)}
\left({\bf x} - {\bf x}' \right) \;f\left({\bf x}' \right)\;d{\bf
x}' = (-1)^{l+m+n} \frac{\partial^{l+m+n}f({\bf x})}{\partial
x_{1}^{l}
\partial x_{2}^{m} \partial x_{3}^{n}},
\end{equation}
where we set ${\bf x} = \left(x_{1},x_{2},x_{3} \right)$.

Using the weak nonlocal distribution~(\ref{weak_distribution}),
one can rearrange the expression on the nonlocal contribution
${\bf D}_{N}$ as follows
\begin{align}
\label{DN_weak} &{\bf D}_{N} = r^{(0)}({\bf x}) N \left(I ({\bf
x}) \right) {\bf E}({\bf x}) + r^{(1)}_{l}\left({\bf x} \right)
\partial_{x_{l}} \left(N\left(I({\bf x}) \right) \; {\bf E}({\bf x})
\right) \nonumber \\
&+ r^{(2)}_{lm} \partial_{x_{l}}\partial_{x_{m}} \left(N \left(
I({\bf x}) \right) \; {\bf E}({\bf x}) \right) + r^{(3)}_{lmn}
\left({\bf x} \right) \partial_{x_{l}}
\partial_{x_{m}}
\partial_{x_{n}} \left( N\left(I ({\bf x}) \right) \; {\bf E}({\bf x})
\right) + \dots
\end{align}
In principle the coefficients $r^{(n)}$ might depend on the
frequency. In particular, we assume the following power dependence
\begin{equation}
r^{(n)} = \frac{\tilde{r}^{(n)}}{\omega^{n}}.
\end{equation}
We consider a model where the function $N$ depends on the
intensity according to the formula
\begin{equation}
\label{Nepsilon} N \left(I(\bf x) \right) = \frac{\varepsilon
\left(I({\bf x}) \right)}{\omega^{\alpha}}
\end{equation}
where $\alpha$ is a real positive constant parameter.

Under the assumptions mentioned above we perform the geometric
optics ({\em semi-classical}) limit of
equation~(\ref{wave_particular}). As usual, let us represent the
electric field in term of the phase $\tilde{S}$
\begin{equation}
\label{E_phase} {\bf E} \left({\bf x} \right) = {\bf E}_{0} \left(
{\bf x}\right) \; e^{i \omega \tilde{S}\left({\bf x} \right)}.
\end{equation}
In high frequency limit, $\omega^{-\alpha}$ is the small parameter
with respect to which we may consider the following asymptotic
expansions
\begin{align}
\label{I_expansion}
&I = I_{0} + \omega^{-\alpha} I_{1} + O \left(\omega^{-2\alpha} \right) \\
\label{epsilon_expansion} &\varepsilon \left(I ({\bf x}) \right) =
\varepsilon_{0} \left(I_{0} ({\bf x}) \right) +  \omega^{-\alpha}
\varepsilon_{1} \left(I_{0}({\bf x}),I_{1}({\bf x}) \right) +
O\left(\omega^{-2\alpha} \right).
\end{align}
 Evaluating ${\bf D}_{N}$  in the limit of high frequency,
one gets
\begin{align}
\label{DN_infinity} {\bf D}_{N} \approx
\varepsilon_{0}\left(I_{0}({\bf x}) \right) {\bf E}_{0}({\bf x})
\; {\cal D} \;e^{i \omega \tilde{S} ({\bf x})}, \quad{} \omega \to
\infty
\end{align}
where
\begin{align*}
&{\cal D} = \tilde{r}^{(0)}  + i \tilde{r}^{(1)}_{l} \;
\tilde{S}_{x_{l}} - \tilde{r}^{(2)}_{lm} \; \tilde{S}_{x_{l}}
\tilde{S}_{x_{m}} - i \tilde{r}^{(3)}_{lmn} \tilde{S}_{x_{l}}
\tilde{S}_{x_{m}} \tilde{S}_{x_{n}}+ \dots, \\ & &l,m,n = 1,2,3.
\end{align*}
Using the expression~(\ref{DN_infinity}), we get
\begin{subequations}
\begin{align}
\label{DN_nabla} \nabla \cdot {\bf D}_{N} &\approx i \omega \;
{\cal D}\; \varepsilon_{0} \left(I_{0} ({\bf x}) \right) \; {\bf
E}_{0} ({\bf x}) \cdot
\nabla \tilde{S} \; e^{i \omega \tilde{S} ({\bf x})} \\
\label{DN_nabla_nabla} \nabla \left(\nabla \cdot {\bf D}_{N}
\right) &\approx - \omega^{2} \;{\cal D}\; \varepsilon_{0}
\left(I_{0} ({\bf x}) \right) \left({\bf E}_{0} \cdot \nabla
\tilde{S} \right) \; \nabla \tilde{S}\; e^{i \omega \tilde{S}
({\bf x})}.
\end{align}
\end{subequations}
Recall that the vector $\nabla \tilde{S}$ is perpendicular to the
wavefront and provides us with the light rays direction. Imposing
the so-called {\em transversality condition}
\begin{equation}
\label{transversality} {\bf E}_{0} \cdot \nabla \tilde{S} = 0
\end{equation}
we restrict ourselves to the solutions such that the electric
field $\bf E$ is perpendicular to the light rays. In this case the
terms~(\ref{DN_nabla}) and~(\ref{DN_nabla_nabla}) do not
contribute to the high frequency limit of the
equation~(\ref{wave_particular}).

\noindent Moreover, we look for $z-$variable slowly depending
solutions of the following form
\begin{equation}
\label{Stilde} \tilde{S} = k z + S\left(x,y,\tau \right)
\end{equation}
where $\tau := z/\omega^{\alpha} $ is the ``slow variable". The
high frequency limit of the equation~(\ref{wave_particular}) at
the leading order gives
\begin{equation}
S_{x}^{2} + S_{y}^{2} = 4 u \left(I_{0} ({\bf x}) \right)
\end{equation}
where $4 u\left(I_{0}({\bf x}) \right) = \varepsilon_{0} \left(
I_{0} ({\bf x})\right) - k^{2}$. \\
If the parameter $\alpha$ is such that $0 < \alpha < 1$ we get an
intermediate contribution between the pure geometric optics order
and the first correction containing the amplitude of the electric
field. In order to calculate it, we note that
\begin{equation}
\tilde{S}_{z}^{2} = \left(k + \omega^{-\alpha} S_{\tau}
\right)^{2} = k^{2} + 2 k \omega^{-\alpha} S_{\tau} + \omega^{-2
\alpha} S_{\tau}^{2}
\end{equation}
where we kept into account that
\begin{equation}
\der{}{z} = \omega^{-\alpha} \der{}{\tau}.
\end{equation}
The $\omega^{2-\alpha}$ order term in
equation~(\ref{wave_particular}) is
\begin{equation}
2 k S_{\tau} + \tilde{{\cal D}} \left(x,y,\tau, S_{x}, S_{y}
\right) \; \varepsilon_{0} + \varepsilon_{1} = 0
\end{equation}
where $\tilde{{\cal D}}$ is a polynomial in $S_{x}$ and $S_{y}$ of
the form
\begin{align}
\tilde{{\cal D}} &= \tilde{r}^{(0)} + i \tilde{r}^{(1)}_{l}
S_{x_{l}} + i k \tilde{r}^{(1)}_{3}- \tilde{r}^{(2)}_{lm}
S_{x_{l}} S_{x_{m}} - k \tilde{r}^{(2)}_{l3} \nonumber \\
&- k \tilde{r}^{(2)}_{3m} S_{x_{m}} + \dots, \quad{} &l,m = 1,2.
\end{align}
Resuming, we have derived the following system of equations
\begin{subequations}
\label{Phase_generalsystem}
\begin{align}
\label{Phase_generalsystem1}
&S_{x}^{2} + S_{y}^{2} = 4 u \left(I_{0} \right) \\
\label{Phase_generalsystem2} &S_{\tau} = \varphi
\left(S_{x},S_{y},I_{0},I_{1},x,y,\tau \right)
\end{align}
\end{subequations}
where we set
\begin{equation}
\varphi \left(x,y,\tau,S_{x},S_{y} \right) = -
\frac{\varepsilon_{0}\left(I_{0} \right)}{2 k} \tilde{\cal D}
\left(S_{x},S_{y},x,y,\tau \right) - \frac{\varepsilon_{1}\left(
I_{0},I_{1}\right)}{2 k }.
\end{equation}

In the construction above, an important r\^ole is played by the
parameter $\omega^{-\alpha}$ with the condition $0< \alpha < 1$.
It is now natural to ask whether such materials exist in reality.
To provide with positive answer let us consider a medium which
satisfies the following dispersion law
\begin{equation}
\label{ColeCole} \varepsilon = \varepsilon_{0} +
\frac{\tilde{\varepsilon}}{1 + \left(2 i \omega \right)^{\alpha}},
\quad{} 0 < \alpha < 1.
\end{equation}
The formula~(\ref{ColeCole}) is referred to as the Cole-Cole
dispersion law and it has been found experimentally by the Cole
and Cole in 1941~\cite{ColeCole}. It is a phenomenoloigical
modification of the ``classical" Debye law (obtained from the
formula~(\ref{ColeCole}) in the limit case $\alpha = 1$). If
$\varepsilon_{0}$ and $\tilde{\varepsilon}$ depend on the
intensity $I$, the function $\varepsilon_{1}\left(I_{0},I_{1}
\right)$ in the expansion~(\ref{epsilon_expansion}) is
\begin{equation}
\label{epsilon1Cole} \varepsilon_{1} \left(I_{0},I_{1} \right) =
\dertot{\varepsilon_{0}}{I_{0}} \left(I_{0} \right) \; I_{1} +
\tilde{\varepsilon} \left(I_{0} \right).
\end{equation}
Thus, our general model is realizable in the Cole-Cole media.

For sake of simplicity, let us consider a linearly polarized
electric field ${\bf E} = E \; \hat{{\bf e }}$, where $\hat{{\bf
e}}$ is a constant unit vector and evaluate explicitly the
nonlocal term ${\bf D}_{N} = D_{N} \; \hat{{\bf e}}$ in the
one-dimensional case
\begin{equation}
\label{DN_1D} D_{N}(x) = \omega^{-\alpha} \int_{-
\infty}^{+\infty} R \left(x'-x \right) \varepsilon \left(I(x')
\right)\; E \left(x' \right) \; dx'
\end{equation}
If the nonlocal distribution function $R\left(x'-x \right)$ is
narrow one as it happens in the high frequency limit, one can
expand $\varepsilon \left(I(x') \right)$ and $E(x')$ around the
point $x$. So, one gets
\begin{align}
\label{DN_1Dexp} D_{N} = &R_{0} \varepsilon \left(I(x) \right) \;
E(x) + R_{1} \left(\varepsilon'\left(I(x) \right)\; E(x)
+\varepsilon \left(I(x) \right)\; E'(x) \right) + \nonumber \\
&R_{2} \left(\frac{1}{2} \varepsilon''\left(I(x) \right) \; E(x) +
 \varepsilon'\left(I(x) \right) \; E'(x) +
  \frac{1}{2} \varepsilon\left(I(x) \right) \; E''(x)  \right)
  \dots
\end{align}
where $\epsilon'= d\epsilon/dx$ and $\epsilon''=
d^{2}\epsilon/dx^{2}$
\begin{equation*}
R_{n} = \int_{-\infty}^{+\infty} R \left(x'-x \right) \left(
x'-x\right)^{n}\; dx'
\end{equation*}
is referred to as $n-$moment of the function $R$. If
\begin{equation}
\label{moments1D} R_{n} = \frac{r_{n}}{\omega^{n}},
\end{equation}
in high frequency limit, one gets
\begin{equation}
\label{DN_1Dhighfreq} D_{N} \approx \frac{\varepsilon_{0}
E_{0}}{\omega^{\alpha}} \left(r_{0} + i r_{1}  S_{x} - \frac{1}{2}
r_{2}  S_{x}^{2} + \dots \right) \; e^{i \omega S} \quad{} \omega
\to \infty.
\end{equation}
We note that narrower is the nonlocal distribution function $R$
smaller are the higher moments. In this particular case the
coefficients $r_{n}$ are related one to another and they are
expressed in terms of the fundamental quantity $R$. Thus, the
equation of $\omega^{2 - \alpha}$ order is
\begin{equation}
S_{\tau} = - \frac{\varepsilon_{0}}{2 k} \left(r_{0} + i r_{1}
S_{x} - \frac{1}{2} r_{2} S_{x}^{2} + \dots \right) -
\frac{\varepsilon_{1}}{2 k}.
\end{equation}
An example of distribution $R$ whose moments are of the
form~(\ref{moments1D}) is provided by the Gaussian distribution
\begin{equation}
R (x) =  \frac{\omega}{\sqrt{\pi}} \; e^{- \omega^{2} x^{2}}.
\end{equation}

In order to give a complete description of the physical system we
should take into account the conservation law of the Poynting
vector. The Poynting vector in complex representation is defined
as follows~\cite{Jackson}.
\begin{equation}
\label{Poynting_full} {\bf P} = {\bf E} \wedge {\bf B}^{*}.
\end{equation}
In the more general case $\mu \neq 1$, $\bf B$ should be replaced
by ${\bf H}$ in the formula~(\ref{Poynting_full}).

\noindent The conservation law is (see pag. 35 in the
reference~\cite{Born})
\begin{equation}
\label{Poynting_conserv} \nabla \cdot {\bf P} = 0.
\end{equation}
Equation~(\ref{maxwell2}) for time-oscillating fields looks like
as follows
\begin{equation}
\label{maxwell2_stat} \nabla \wedge {\bf E} + i \omega {\bf B} =
0.
\end{equation}
Using the representation
\begin{equation*}
{\bf E} = {\bf E}_{0} \; e^{i \omega \tilde{S}}, \quad{} {\bf B} =
{\bf B}_{0}\; e^{i \omega \tilde{S}},
\end{equation*}
one gets
\begin{equation}
\nabla \wedge {\bf E}_{0} + i \omega \nabla \tilde{S} \wedge {\bf
E}_{0} = i \omega {\bf B}_{0}.
\end{equation}
The leading order in high frequency limit gives
\begin{equation}
\label{B0_highfreq} {\bf B}_{0} = \nabla \tilde{S} \wedge {\bf
E}_{0}.
\end{equation}
Thus, we have the following approximation of the Poynting vector
\begin{align*}
{\bf P} = {\bf E}_{0} \wedge {\bf B}_{0} = {\bf E}_{0} \wedge
\left(\nabla \tilde{S} \wedge {\bf E}_{0}^{*} \right) = \nonumber
\\
\left({\bf E}_{0} \cdot {\bf E}_{0}^{*} \right) \; \nabla
\tilde{S} - \left({\bf E}_{0} \cdot \nabla \tilde{S} \right) \;
{\bf E}_{0}^{*}
\end{align*}
In virtue of the transversality condition~(\ref{transversality})
one finally has
\begin{equation}
\label{Poynting_full2} {\bf P} = I\; \nabla \tilde{S}.
\end{equation}
where $I =\left|{\bf E}_{0} \right|^{2}$. Note that ${\bf P}$ is
parallel to the gradient of the phase $\nabla S$. Using the
expression~(\ref{Poynting_full2}) in~(\ref{Poynting_conserv}), one
obtains
\begin{equation}
\label{Poynting_conserv_I} \nabla I \cdot \nabla \tilde{S} + I
\nabla^{2} \tilde{S} = 0.
\end{equation}
Due to the slow dependence on the variable $z$, for phases of the
form~(\ref{Stilde}) and the asymptotic expansion $I = I_{0} +
\omega^{-\alpha} I_{1} + \dots$,
equation~(\ref{Poynting_conserv_I}) becomes
\begin{subequations}
\label{Intensity_generalsystem}
\begin{align}
\label{Intensity_generalsystem1} &I_{0} \left(S_{xx} + S_{yy}
\right) + I_{0x} S_{x} + I_{0y} S_{y}
= 0 \\
\label{Intensity_generalsystem2} &I_{1} \left (S_{xx} + S_{yy}
\right) + I_{1x} S_{x} + I_{1y} S_{y} + k I_{0\tau} = 0.
\end{align}
\end{subequations}
The system~(\ref{Intensity_generalsystem}) has to be considered
together with the system for the
phase~(\ref{Phase_generalsystem}).

The properties of present model have to be investigated by the
joint analysis of the system~(\ref{Phase_generalsystem}) and the
system~(\ref{Intensity_generalsystem}). In our discussions we will
separate the compatibility analysis of the transverse
equation~(\ref{Phase_generalsystem1})
and~(\ref{Intensity_generalsystem1}) from the ``evolution"
equations~(\ref{Phase_generalsystem2})
and~(\ref{Intensity_generalsystem2}).

\section{Paraxial light beams.}
\label{sec_paraxial}

\subsection{The nonlocal NLS equation.} \label{sec_nonlocal}
Many phenomenological models are based on the so-called nonlocal
NLS
equation~\cite{Snyder,Assanto,Krolik1,KrolikExact,Krolik2,Krolik3,
Moro1,Moro2,MoroSpie}. It describes paraxial light beams in
nonlinear (and also nonlocal) media. In the present section we
discuss the connection between the high frequency model discussed
above and the nonlocal NLS equation.

Let us consider a displacement vector of the form
\begin{equation}
{\bf D} = \sigma {\bf E} + \sigma^{3} {\bf D}^{(3)},
\end{equation}
where $\sigma$ is a constant parameter. The paraxial approximation
in equation~(\ref{wave_general}) is usually performed by the
consideration of the slow variations of a small-amplitude electric
field (or formally by the substitutions)
\begin{equation*}
\left (\partial_{x}, \partial_{y},\partial_{z} \right) \to \left(
\sigma \partial_{x}, \sigma \partial_{y}, \sigma^{2}
\partial_{z}\right), \quad{} {\bf E} \to \sigma {\bf E}
\; e^{i \omega z}.
\end{equation*}
The leading nontrivial order in equation~(\ref{wave_general}) is
\begin{equation}
2 i  \omega \der{{\bf E}}{z} + \nabla_{\bot}^{2} {\bf E} +
\omega^{2} {\bf D}^{(3)} = 0,
\end{equation}
where $\nabla_{\bot} = \left(\partial_{x}, \partial_{y} \right)$.
For a general nonlocal nonlinear medium we can write
\begin{equation}
\label{D3def} {\bf D}^{(3)} = \int_{\mathbb{R}^{3}} R\left({\bf
r'} - {\bf r}; a \right) \; N\left(I({\bf r'}) \right) {\bf
E}({\bf r'}) \; d^{3}{\bf r'}
\end{equation}
where ${\bf r} = \left(x,y,z \right)$. We use the notation ${\bf
D}^{(3)}$ to recall that, in the case of a local Kerr medium, the
relation~(\ref{D3def}) is reduced to the cubic nonlinearity which
is associated with the standard NLS equation~\cite{Boyd}. The
distribution $R({\bf r}'-{\bf r};a)$ characterizes the nonlocal
response around the point ${\bf r}$ and $a$ is the ``width"
parameter (in the following it will be assumed to be depending on
the frequency $\omega$). $N(I)$ is a certain nonlinear response
depending on the intensity of the electric field $I = \left |{\bf
E} \right|^{2}$.

We note that due to the paraxial approximation the nonlocal
response along $z$ can be neglected. As illustrative example to
explain this fact let us consider a Gaussian nonlocal response
\begin{equation}
R\left({\bf r}',{\bf r} \right) = \frac{1}{\left(2 \pi
\right)^{\frac{3}{2}}} \; e^{- \left|{\bf r}' - {\bf r} \right
|^{2}}
\end{equation}
where $\left|{\bf r} \right|^{2} = x^{2} + y^{2} + z^{2}$. Due to
the paraxial approximation we take into account of slow dependence
on the variable $z$ by the substitution $z \to \epsilon^{-1} z$
\begin{equation}
\label{paraxial_R} \lim_{\epsilon \to 0} \frac{1}{\epsilon (2
\pi)^{\frac{3}{2}}} \; e^{\frac{(z'-z)^{2}}{\epsilon^{2}}} \;
e^{(x'-x)^{2}+ (y'-y)^{2}} = \frac{1}{2 \pi} \; \delta\left(z'-z
\right)\; e^{(x'-x)^{2}+(y'-y)^{2}}.
\end{equation}
The Dirac $\delta-$function in the left hand side
of~(\ref{paraxial_R}) implies that the response along the
direction $z$ becomes local. Thus, with the use of the
distribution~(\ref{paraxial_R}) in the general definition of ${\bf
D}^{(3)}$ the $3$D integral is reduced to a $2-$dimensional one
\begin{equation}
\label{D3def2D} {\bf D}^{(3)} = \int_{-\infty}^{+\infty}
\int_{-\infty}^{+\infty} R\left(x'-x,y'-y,a \right)\;
N\left(I(x,y,z) \right)\; {\bf E}(x,y,z)\; dx\; dy.
\end{equation}
The model considered above, can be further simplified under
suitable assumptions on the widths of the nonlocal distribution
$R$ and the nonlinear response $N(I)$.
 For sake of simplicity we focus on the $1+1-$dimensional case
\begin{equation}
\label{1DNNLS_general} 2 i \omega \der{{\bf E}}{z} + \dersec{{\bf
E}}{x} + \omega^{2} {\bf D}^{(3)} =  0
\end{equation}
where
\begin{equation}
\label{constitutive} {\bf D}^{(3)} = \int_{-\infty}^{+\infty}
R(x-x';a) \; N \left(I(x') \right ) \; {\bf E}(x')  dx'.
\end{equation}
Let us define the width $\delta R$ of the nonlocal distribution
$R\left (x-x';a \right)$  as the minimum such that
\begin{equation}
R \left (x-x';a \right) \simeq 0, \quad{} \forall x' \notin \left
[ x- \delta R, x+ \delta R\right ].
\end{equation}
Of course, $\delta R$ depends on the width parameter $a$.
Similarly, we can introduce the widths $\delta E$ and $\delta N$
of the electric field and the nonlinear response respectively.
Suppose they satisfy the following conditions
\begin{equation}
\label{wideness} \delta R \sim \delta N, \quad{} \delta R <<
\delta E.
\end{equation}
Moreover, we assume the nonlinear response $N(I)$ to be of the
form
\begin{equation}
\label{N_gamma} N \left (I (x) \right) = \tilde{N} \left (X
\right),
\end{equation}
where $X = \gamma \: x$ and $\gamma \propto  1/ \delta N$.
Expanding ${\bf E}(x')$ and $N\left (I(x') \right )$ in Taylor
series around $x$, one gets the following approximation of the
formula~(\ref{constitutive})
\begin{equation}
\label{bang_model} {\bf D}^{(3)} \simeq \left ( \int_{- \infty}^{+
\infty} R \left (x-x';a \right) N \left (I(x') \right) dx' \right
) {\bf E}(x),
\end{equation}
where we kept into account that due to equation~(\ref{N_gamma})
higher orders of the expansion of $N\left (I(x) \right)$ are not
negligible. Note that the formula~(\ref{bang_model}), in the case
of nonlocal Kerr-type medium, leads to the nonlocal nonlinear
Schr\"odinger equation discussed in the paper~\cite{KrolikExact}.

\noindent For instance, given a bell-shape electric field ${\bf E}
= {\bf E}_{0} \; \exp \left [- x^{2}/2 \sigma^{2} \right]$, a
nonlinear response of the form $N(I) = I^{\alpha} = \left |{\bf
E}_{0} \right |^{2 \alpha} \; \exp \left [- \left (\gamma x
\right)^{2}/{2 \sigma^{2}} \right]$, where $\gamma = \sqrt{2
\alpha}$, satisfies the condition~(\ref{N_gamma}). Nevertheless,
it is easy to see that only the validity of
relations~(\ref{wideness}) is sufficient to obtain the
model~(\ref{bang_model}). For instance, for ${\bf E} = {\bf
E}_{0}/ \cosh^{2}(x)$  and $N \left (I \right ) = I^{\alpha}$,
condition~(\ref{wideness}) is verified for $\alpha$ large enough.
Finally, we note that it is straightforward to generalize the
previous considerations to construct a more general $(2+1)$D
model.

\subsection{High frequency regimes.}
\label{sec_integrable} Now, we discuss the above in high frequency
regime.

\noindent With the representation
\begin{equation*}
{\bf E} = {\bf E}_{0} \; e^{i \omega \tilde{S}(x,y,z)},
\end{equation*}
 the NNLS equation takes the form
\begin{equation}
\label{phase_NNLS} 2 \tilde{S}_{z} + \left(\nabla_{\bot} \tilde{S}
\right)^{2} = N_{0}\left(I_{0} \left(x,y,z \right) \right),
\end{equation}
where $\nabla_{\bot} =\left(\partial_{x},\partial_{y} \right)$ and
$N_{0}(I_{0})$ is the high frequency limit of the intensity law
\begin{equation*}
\simbon{\lim}{\omega \to \infty} N \left(I \right) = N_{0}
\left(I_{0} \right)
\end{equation*}
Paraxial approximation of the Poynting vector conservation
law~(\ref{Poynting_conserv}) at the leading order on $\sigma$
gives the following equation on the $xy-$plane
\begin{equation}
\label{phase_constraint} \nabla_{\bot} I \cdot  \nabla_{\bot}
\tilde{S} + I \nabla_{\bot}^{2} \tilde{S} =0.
\end{equation}
An analysis of the $(1+1)$D reduction of the couple of the
equations~(\ref{phase_NNLS}) and~(\ref{phase_constraint}) suggests
that possible stable light beams (in this specific regime) exists
only in $(2+1)D$. Indeed, the following example shows that the
bell-shape initial beam profiles are no longer preserved.

\noindent If $S$ in equations~(\ref{phase_NNLS})
and~(\ref{phase_constraint}) does not depend on the variable $y$,
one has
\begin{subequations}
\label{1D_high}
\begin{align}
\label{1D_phase}
&2 \tilde{S}_{z} + \tilde{S}_{x}^{2} = N_{0}\left(I_{0} \right) \\
\label{1D_conservation} &I_{0} \tilde{S}_{xx} + I_{0x}
\tilde{S}_{x} = 0.
\end{align}
\end{subequations}
Integrating equation~(\ref{1D_conservation}), one gets
\begin{subequations}
\label{phase_1D}
\begin{align}
\tilde{S}_{x} &= \frac{\gamma_{0}}{I_{0}} \\
\tilde{S}_{z} &= \frac{1}{2} \left(N_{0}(I_{0}) -
\frac{\gamma_{0}^{2}}{I_{0}^{2}} \right),
\end{align}
\end{subequations}
where $\gamma_{0}$ is an arbitrary real constant. In virtue of
$S_{xz} = S_{zx}$ one obtains
\begin{equation}
\label{compatibility_1D} I_{0z} + \left(\frac{I_{0}^{2}
N_{0}'(I_{0})}{2\gamma_{0}} + \frac{\gamma_{0}}{I_{0}} \right)
I_{0x} = 0,
\end{equation}
where $N'_{0} := \dertot{N_{0}}{I_{0}}$. General solution of
equation~(\ref{compatibility_1D}), calculated by the
characteristics method~\cite{Courant}, is provided by the
following implicit relation
\begin{equation}
\label{characteristic} x - \left(\frac{I_{0}^{2} N'(I_{0})}{2
\gamma_{0}} + \frac{\gamma_{0}}{I_{0}} \right) - \Phi \left(I_{0}
\right) = 0
\end{equation}
where $\Phi$ is an arbitrary function of its argument. It is
assigned by the initial profile of the intensity
\begin{equation}
\label{initial_1D} I_{0}\left(x,0 \right) = \Phi^{-1}\left(x
\right).
\end{equation}
It is well known~\cite{whithambook} that equations of
type~(\ref{compatibility_1D}) exhibit breaking wave phenomena for
finite $z$. For example, a smooth initial profile, such that
$I_{0} \to 0$ for $x \to \pm \infty$ breaks at finite $z$ where
$I_{0x} \to \infty$.

Let us now consider a $(2+1)$D model for the following class of
solutions
\begin{equation}
\label{phase_const_z} \tilde{S} = k z + S \left(x,y,\tau \right),
\end{equation}
where $k$ is a real constant and $\tau = \omega^{\alpha} z$ (with
$\alpha > 0$). A class of nonlocal perturbations with respect to
the slow variable $\tau$ will be discussed in the following
subsection. Under these assumptions equations~({\ref{phase_NNLS}})
and~(\ref{phase_constraint}) assume the following form
\begin{subequations}
\label{transverse_system}
\begin{align}
\label{transverse_system_eikonal}
&S_{x}^{2} + S_{y}^{2} = 4 u\left(I_{0}(x,y) \right) \\
\label{transverse_system_conservation} &I_{0} \left(S_{xx} +
S_{yy} \right) + I_{0x} S_{x} + I_{0y} S_{y} = 0
\end{align}
\end{subequations}
where $4 u(I_{0}) = \tilde{N}_{0}(I_{0}) - 2 k$. We kept into
account that due to the rule
\begin{equation*}
\der{}{z} = \omega^{-\alpha} \der{}{\tau}
\end{equation*}
$z-$derivatives do not contribute in the limit $\omega \to
\infty$. We note also that $I_{0}$ is the leading order term of
the $\omega^{-\alpha}-$asymptotic expansion of the intensity
\begin{equation*}
I(x,y,z) = I_{0} (x,y,\tau) + \omega^{-\alpha} I_{1}(x,y,\tau) +
\dots\; .
\end{equation*}
One refers to the functional dependence among $u$ and $I_{0}$ as
{\em intensity law}. It is determined by the specific physical
properties of the medium. Note that once $u(I_{0})$ is given, the
system~(\ref{transverse_system}) is overdetermined. Its
consistency condition will be discussed in the next section.

\subsection{Nonlocal perturbations.}
\label{sec_nonlocal_perturbation} Let us consider a model for
which the nonlocal nonlinear contribution to the displacement
vector is of the following form
\begin{equation}
\label{dvn_displacement} {\bf D}^{(3)} = \varepsilon {\bf E} +
{\bf D}_{N}
\end{equation}
where
\begin{equation*}
{\bf D}_{N} = \int_{\mathbb{R}^{2}} \; R\left({\bf r}'- {\bf r}
,\omega\right) \; N\left(I({\bf r}',z)\right)  {\bf E}({\bf r},z)
d{\bf r}'
\end{equation*}
In particular, we are interested in the construction of a special
class of weak nonlocal perturbations given by a nonlocal
distribution function of the form
\begin{equation}
\label{dirac_nonlocal} R\left({\bf r} -{\bf r}',\omega \right) =
\sum_{l,m=0}^{M} R_{l,m}\left({\bf r},\omega \right) \;
\delta^{(l,m)} \left({\bf r} -{\bf r}' \right),
\end{equation}
where ${\bf r} = \left( x,y\right)$ the distributions
$\delta^{(l,m)}$ are defined by~(\ref{dirac_def}) and $M$ is an
arbitrary integer referred to as {\em nonlocality degree}. As a
consequence one gets
\begin{equation}
\label{fnloc} {\bf D}_{N} = \sum_{l=1}^{2} c_{n} \der{{\bf
E}}{x_{l}} + \sum_{l,m=1}^{2} c_{l,m} \dermixd{{\bf
E}}{x_{l}}{x_{m}} + \dots\;\;\;.
\end{equation}
Using the following asymptotic expansions on $\omega^{-\alpha}$
\begin{gather}
\label{expansion}
\begin{aligned}
{\bf E}_{0}(x,y,z) &= {\bf E}_{0}(x,y,\tau) + \omega^{-\alpha}
{\bf E}_{1} (x,y,\tau) + \dots \\
\varepsilon(x,y,z) &= \varepsilon_{0}(x,y,\tau) + \omega^{-\alpha}
\varepsilon_{1}^{2} (x,y,\tau) + \dots \;\;,
\end{aligned}
\end{gather}
and proceeding similarly to the previous section for a Cole-Cole
medium~(\ref{ColeCole}), one gets from the NNLS equation in the
leading orders $\omega^{2}$ and $\omega^{2-\alpha}$ the following
system
\begin{subequations}
\label{phase}
\begin{align}
\label{eikonal}
&S_{x}^{2} + S_{y}^{2} = 4 u, \\
&\label{z_dependence} S_{\tau} = \varphi(S_{x},S_{y},x,y,\tau)
\end{align}
\end{subequations}
where $4 u = \varepsilon_{0} - 2 k$ and the function $\varphi$ is
a polynomial in $S_{x}$ and $S_{y}$. Equation~(\ref{eikonal}) is
the well known eikonal equation in two-dimensions. The function
$\varphi$ in equation~(\ref{z_dependence}), in virtue of
expression~(\ref{fnloc}), is an $M$-degree polynomial in $S_{x}$
and $S_{y}$ and it describes an $M$-degree nonlocal response.
Moreover, if we require that equation~(\ref{z_dependence})
possesses the phase inversion symmetry ($S \to - S$) just like the
eikonal equation, the function $\varphi$ must contain only odd
degree terms in $S_{x}$ and $S_{y}$. We note also that the system
of equations~(\ref{phase}) has been derived first directly from
the Maxwell equations~\cite{Moro1,Moro2}.

In local case we have
\begin{equation}
\label{local_hp} \varphi = \varphi (z), \quad{} u  = u (x,y),
\end{equation}
where $\varphi$ and $u$ are certain function of their arguments.
For a first degree nonlocality we have
\begin{equation}
\label{firstnonlocal_phi_hp} \varphi = \alpha_{1} S_{x} +
\alpha_{2} S_{y}, \quad{}
\end{equation}
and $u$ must satisfy the following linear equation
\begin{equation}
\label{firstnonlocal_u_hp} u_{z} = \left (\alpha_{1} u\right)_{x}
+ \left (\alpha_{2} u\right)_{y},
\end{equation}
where $\alpha_{1}$ and $\alpha_{2}$ are harmonic conjugate
functions, that is they satisfy the Cauchy-Riemann conditions
$\alpha_{1x}$ $=$ $\alpha_{2y}$ $\alpha_{1y}$ $=$ $- \alpha_{2x}$.

\section{Transverse equations.}
\label{sec_transverse}
 The compatibility of the system of equations for the
phase~(\ref{Phase_generalsystem}) and
intensity~(\ref{Intensity_generalsystem}) results to be highly
nontrivial problem. More specifically, one can see that it is not
consistent for arbitrary intensity law. The construction presented
above suggest to separate the compatibility analysis on the
$xy-$plane from the evolution along the $z-$direction. We will
study the system of equations~(\ref{Phase_generalsystem1})
and~(\ref{Intensity_generalsystem1}) for a general form of the
intensity law and then we will discuss the ``evolution" of the
solutions with respect to the variable $\tau$.

\subsection{Elliptic intensity laws.}
\label{sec_elliptic} Let us assume the function $u(I_{0})$ to be
invertible. Note that for various physically meaningful models
such as the Kerr-type media
\begin{equation}
\label{Kerr_law} u = I_{0}^{\gamma}, \quad{} \gamma > 0
\end{equation}
and logarithmic saturable media
\begin{equation}
\label{log_law} u = \log \left(1 + I_{0}/I_{0t}\right)
\end{equation}
where the constant $I_{0t}$ is the so-called threshold intensity,
intensity law is the invertible one.

Then, we can consider the function $I_{0} = I_{0}(u)$. We refer to
it as {\em inverse intensity law}. Since
\begin{equation*}
\nabla_{\bot} I_{0} = I_{0}' \nabla_{\bot} u
\end{equation*}
 where $I_{0}' = dI_{0}/du$, the system of
 equations~(\ref{Phase_generalsystem1})and~(\ref{Intensity_generalsystem1})
(or~(\ref{transverse_system})) looks like as follows
\begin{subequations}
\label{transverse_system_part}
\begin{align}
\label{transverse_system_part1}
&S_{x}^{2}  + S_{y}^{2} = 4 u \\
\label{transverse_system_part2} &I_{0} \left(S_{xx} + S_{yy}
\right) + I_{0}' \left(u_{x} S_{x} + u_{y} S_{y} \right) = 0.
\end{align}
\end{subequations}
Substituting the expression for $u$
from~(\ref{transverse_system_part1})
into~(\ref{transverse_system_part2}), we obtain the following
second order partial differential equation
\begin{equation}
\label{S_elliptic} A S_{xx} + B S_{yy} + 2 C S_{xy} = 0,
\end{equation}
where by definition $J_{0} = \log I_{0}$ and
\begin{equation}
A = J_{0}\;' S_{x}^{2} + 2, \quad{} B = J_{0}\;' S_{y}^{2} + 2,
\quad{} C = J_{0}\;' S_{x} S_{y}.
\end{equation}
The equations of the form~(\ref{S_elliptic}) are well studied in
mathematics (see e.g.~\cite{Bojarski} and references therein).
Their properties are critically depending on the signature of the
discriminant
\begin{equation}
\label{delta_def} \Delta = A B - C^{2}.
\end{equation}
In particular, one can distinguish three cases, elliptic
($\Delta>0$), parabolic ($\Delta = 0$) and hyperbolic ($\Delta <
0$). In what follows, we will focus on the elliptic case motivated
by the fact that the intensity law for Kerr-type~(\ref{Kerr_law})
and logarithmic saturable media~(\ref{log_law}) satisfy the
ellipticity condition $\Delta>0$ uniformly. We remark that these
models are very relevant in practical applications since they are
associated with the propagation of stable spatial
solitons~\cite{Sulem,saturable}

We would like to emphasize that elliptic second order nonlinear
equations of the form~(\ref{S_elliptic}) possess several
remarkable analytical and geometrical properties (see
e.g.~\cite{Bojarski,Iwaniec,Bers,Vekua,Ahlfors}). An interesting
class of solutions of equation~(\ref{S_elliptic}) is provided by
the {\em Beltrami equation} which is well known and well studied
in the theory of elliptic systems of PDEs and in the theory of
{\em quasiconformal mappings}\cite{Vekua,Ahlfors}. Let us
introduce the complex variable $z = x + i y$ (it should not be
confused with real variable associated with the $z-$axis) and the
complex gradient $w = S_{x} - i S_{y}$. In these notations
equation~(\ref{S_elliptic}) takes the form
\begin{equation}
\label{S_elliptic_complex} a w_{z} + b w_{\bar{z}} + \bar{a}
\bar{w}_{\bar{z}} + \bar{b} \bar{w}_{z} = 0,
\end{equation}
where
\begin{align}
a = \frac{1}{2} \left(A - B + 2 i C \right), \quad{} b =
\frac{1}{2} \left (A + B - 2 i C \right).
\end{align}
It is immediate to verify that if $w$ satifies the, so-called,
nonlinear Beltrami equation
\begin{equation}
\label{Beltrami_nlin} w_{\bar{z}} = \mu \left (w, \bar{w} \right)
w_{z}, \quad{} \mu = - \frac{a}{b}.
\end{equation}
then it is also a solution of equation~(\ref{S_elliptic_complex}).
It is easy to verify that ellipticity condition $\Delta
> 0$ implies $\left | \mu \right | < 1$.
Function $\mu$ possesses a remarkable geometrical meaning being
the {\em complex dilatation} of the quasiconformal mapping $w$.

Thus, the evolution of a light beam profile along the direction
$\tau$ turns out to be described by suitable deformations of
quasiconformal mappings.

More specifically, let us consider an input light beam profile
such that $I_{0} \neq 0 $ inside a simply connected domain $G$ and
$I_{0} = 0$ onto the smooth boundary $\Gamma$ of $G$ and outside
it. Set $u_{0} = u(I_{0} = 0)$. Writing down the eikonal
equation~(\ref{transverse_system_eikonal}) in terms of $w$ and
$z$, one gets
\begin{equation}
\label{w_eikonal} w (z) \bar{w} (z) = 4 u.
\end{equation}
From equation~(\ref{w_eikonal}) one gets $\left |w \right| = 2
\sqrt{u_{0}}$ for $z \in \Gamma$, that is $\Gamma$ is mapped on
the $2 \sqrt{u_{0}}$-radius circle. Assuming $w$ to be assigned in
such a way that, for instance, $w(0) = 0$, where $0 \in G$, and
the variation of the argument of the complex number $w$ around
$\Gamma$ is $\Delta_{\Gamma} \arg w = 2 \pi$, it can be proven
that $w$ is a homeomorphic mapping of domain $G$ onto
$2\sqrt{u_{0}}-$radius disk $\Gamma\;'$~\cite{Iwaniec}. Of course,
mapping $w$ preserves the topology of domain $G$. In virtue of the
assumption~(\ref{phase_const_z}) the wavefront evolves according
to the equation
\begin{equation}
\omega^{-\alpha} k \tau + S(x,y,\tau) = \textup{const}.
\end{equation}

\begin{figure}[h]
\centerline{\includegraphics[width=8cm]{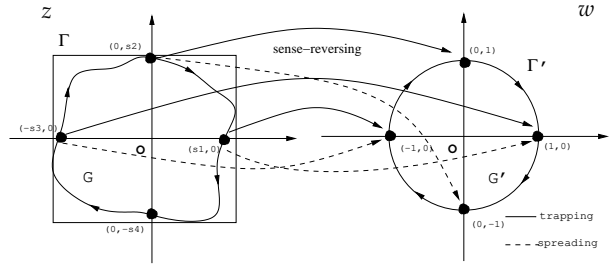}} \caption{
\small{Sense-reversing mappings (solid line) describe beams which
tend to be confined inside the rectangle or scattered (dashed
line) outside it.} \label{nbeltrami_a}}
\end{figure}

\begin{figure}
\centerline{\includegraphics[width=8cm]{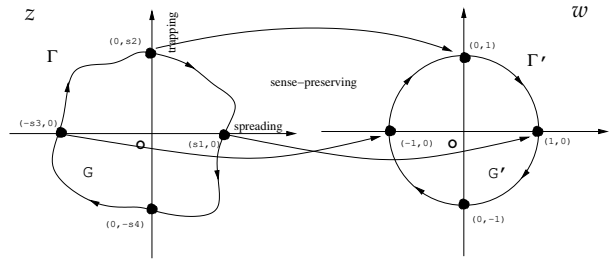}}
\caption{\small{For the sense-preserving mapping the light beam
spreads along $x-$direction and tends to be trapped along
$y-$direction.} \label{nbeltrami_b}}
\end{figure}

We observe that the mapping $w = S_{x} - i S_{y} = \left(S_{x}, -
S_{y} \right)$ can be also regarded as a two-dimensional vector
field on the $z-$plane associated with transverse components of
the wavefront normal unit vector
\begin{equation}
\label{unit_vec} {\bf n} = \frac{\nabla \tilde{S}}{|\nabla
\tilde{S} |} = \left(\frac{S_{x}}{\sqrt{k^{2}+ 4 u}},
\frac{S_{y}}{\sqrt{k^{2} + 4 u}}, \frac{1}{\sqrt{k^{2}+4u}}
\right).
\end{equation}
We remind that the vector ${\bf n}$ is, by definition, parallel to
the Poynting vector.
 This means that the mapping $w$ encodes information
about light-rays distribution around direction $\tau$. For
example, let us consider the mapping given in
figure~\ref{nbeltrami_a}. Curves $\Gamma$ and $\Gamma'$ are
oriented leaving domain on the right hand side. Under the
assumptions mentioned above, there exists a homeomorphism $w$ of
domain $G$ onto $G'$ acting in such a way that $w(s_{1},0) =
(-1,0)$, $w(0,s_{2}) = (0,1)$ and $w(-s_{3},0) = (1,0)$.
Consequently, the normal unit vector to wave-front on the boundary
$\Gamma$ is oriented in such a way that the light rays lie inside
the rectangle circumscribing the domain $G$ (see
figure~\ref{nbeltrami_a}). Recall that $y$-component of $w$
reverse $y$-component of $\bf{n}$. Then, this mapping describes a
light beam `trapped' around the direction $\tau$. Conversely, if
$w(s_{1},0) = (1,0)$, $w(0,s_{2}) = (0,-1)$ and $w(-s_{3},0) =
(0,-1)$ radiation spreads transversely far from $\tau-$direction.
Note also that in both cases homeomorphism $w$ is sense-reversing.
If we consider a sense-preserving mapping such as, for instance,
$w(s_{1},0) = (1,0)$, $w(0,s_{2}) = (0,1)$ and $w(-s_{3},0) =
(-1,0)$, the beam spreads along $x-$direction and tends to be
trapped along $y-$axis (see figure~\ref{nbeltrami_b}). The
boundary conditions and the orientation of the mapping allows us
to control strongly the properties of the light beam.

We emphasize, finally, that all of these observations can be
generalized to the case of arbitrary $n-$connected domains. They
could be used for description of interacting light beams.

\noindent Another class of solutions of
equation~(\ref{S_elliptic_complex}) can be obtained solving the
equation
\begin{equation}
\label{pre_reciprocal} b w_{\bar{z}} + \bar{a} \bar{w}_{\bar{z}} =
0.
\end{equation}
Introducing the reciprocal coordinates by inversion of the system
\begin{gather}
\label{recip_coordinates}
\begin{aligned}
z &= z(w,\bar{w}) \\
\bar{z} &= \bar{z} (w,\bar{w}),
\end{aligned}
\end{gather}
one converts equation~(\ref{pre_reciprocal}) into the form of the
{\em linear} Beltrami equation
\begin{equation}
\label{Beltrami_lin} z_{\bar{w}} = \nu(w,\bar{w}) z_{w},
\end{equation}
where $\nu (w,\bar{w}) = - \bar{a}/b$ and, due to ellipticity,
$|\nu| < 1$. The advantage of consideration of
equation~(\ref{pre_reciprocal}) in reciprocal coordinates is that
the Beltrami equation is linear and it can be solved explicitly
for different choices of the intensity law. Moreover, it is well
known that several theorems in analytic functions theory can be
rigorously generalized to the quasi-analytic functions which are
the solutions of equation~(\ref{Beltrami_lin}) (see e.g.
Ref.~\cite{Vekua}). In particular, a generalization of so-called
Liouville theorem ({\em Vekua's theorem}) holds. If $z =
z(w,\bar{w})$ is bounded on whole $w-$plane and satisfies the
linear Beltrami equation~(\ref{Beltrami_lin}) it can be shown that
$z (w,\bar{w}) \equiv \textup{constant}$~\cite{VekuaTh}. We stress
that for the constant solution, mapping from $w-$to$-z-$plane is
singular and reciprocal transformation~(\ref{recip_coordinates})
is not defined. Then, any non-trivial solution of
equation~(\ref{Beltrami_lin}) must be singular somewhere on the
complex plane and different type of singularities can occur, such
as poles, essential singularities, singularities of the
derivatives etc.
\begin{figure}
\centerline{\includegraphics[width=8cm]{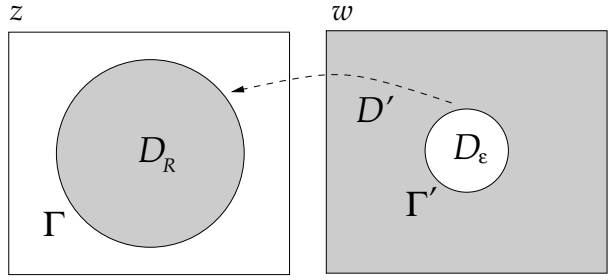}}
\caption{\small{Homeomorphic mapping from the exterior of
$D_{\epsilon}$ on $D_{R}$ such that $\Gamma'$ goes onto $\Gamma$.
Light rays associated with the points outside $D_{R}$ are parallel
to the $z-$axis as much as $\epsilon$ is small.}
\label{beltrami_pole}}
\end{figure}
As illustrative example, we focus our attention on a solution
$z(w,\bar{w})$ which have a simple pole at $w = 0$. As discussed
above, one can always consider a homeomorphism from $D' =
\mathbb{C} \setminus D_{\epsilon}$ to $D_{R}$, where
$D_{\epsilon}$ is a disk of arbitrarily small radius $\epsilon$ on
the $w-$plane and $D_{R}$ is the $R$-radius disk on the $z-$plane,
mapping the boundary of $D_{\epsilon}$ on the boundary of $D_{R}$.
Inverse mapping $w : \; D_{R} \to D'$, constructed in such a way,
can be used to describe a beam ``confined''around $\tau-$axis.
Indeed, the transverse component of the vector ${\bf n}$ outside
$D_{R}$ are arbitrarily small. Hence, the light rays can be
settled down parallel to the $\tau-$axis with arbitrary accuracy.
So, the beam results to be self-guided around $\tau$ as much as
disk $D_{\epsilon}$ is small.

Coming back to nonlinear Beltrami equation~(\ref{Beltrami_nlin}),
we expect that for ``mild enough'' complex dilatations $\mu$,
Vekua's theorem still holds. In these cases, the only one bounded
solution on whole $z-$plane is $w = \textup{constant}$. In many
physical situations, the intensity distribution on the $z-$plane,
at certain $\tau$, goes to zero for $z \to \infty$, or
equivalently, one can says that intensity vanishes outside a big
enough $R-$radius disk $D_{R}$. Thus, outside $D_{R}$, refractive
index assumes a constant value $u = u_{0}$ and the solutions of
eikonal equation~(\ref{transverse_system_eikonal}) is
\begin{equation}
S = c_{0} x + c_{1} y + c_{3},
\end{equation}
where $c_{0}$, $c_{1}$ and $c_{3}$ are constants and the condition
$c_{0}^{2} + c_{1}^{2} = 4 u_{0}$ holds. For a paraxial beam we
have $c_{0} = c_{1} = 0$. As the consequence $w = S_{x} - i S_{y}
= c_{0} - i c_{1} = 0$ and, of course it satisfies Beltrami
equation in $\mathbb{C} \setminus D_{R}$. In virtue of Vekua's
theorem, the only one bounded solution is $w \equiv 0$. Then, any
non-trivial solution must be singular somewhere on the plane. In
our example, wavefront is approximately plane for $z \in
\mathbb{C} \setminus D_{R}$ and possesses singularity inside
$D_{R}$.

In the next subsection we will discuss the so-called optical
vortices obtained from equation~(\ref{S_elliptic}) assuming the
phase to be a harmonic function on the $xy-$plane. We will see
that, in this case, equation~(\ref{S_elliptic}) is equivalent to
the minimal surfaces equation. We refer to this class of solutions
as {\em minimal sector} in order to distinguish it from the {\em
Beltrami sector} which is associated with the solutions of
equations~(\ref{Beltrami_nlin}) and~(\ref{Beltrami_lin}).

\subsection{Helicoidal wavefronts.}
\label{sec_vortex} The successive approximations method is a
general approach to solve the linear Beltrami equation (it has
been demonstrated by Tricomi~\cite{Tricomi}). Nevertheless,
calculation of exact `explicit' solutions can be the difficult
task and the chances of success strongly depend on the form of the
complex dilatation. Incidentally, we note that solutions
possessing cylindrical symmetry such that
\begin{equation}
\label{cylindrical_sol} S = S(r), \quad{} u = u(r), \quad{} r =
\sqrt{x^{2} + y^{2}},
\end{equation}
are not compatible with intensity law. It is straightforward to
verify that equations~(\ref{transverse_system}) along with
assumptions~(\ref{cylindrical_sol}) imply that intensity depends
explicitly on $z-$axis distance $r = \sqrt{x_{2}+y^{2}}$
\begin{equation}
\label{cylindrical_int} I_{0} = \frac{1}{2 c r \sqrt{u}},
\end{equation}
where $c$ is an arbitrary constant.

\begin{figure}[h]
\centerline{\includegraphics[width=12cm]{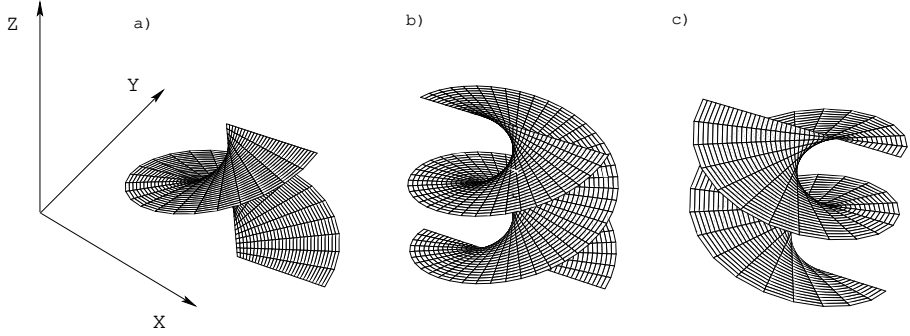}}
\caption{Helicoidal structures of wavefront around $\tau-$axis: a)
one-start right-screw ($K=1$); b) two-start right screw ($K=1$);
c) two-start left screw ($K=-1$).} \label{figure2}
\end{figure}

Here, we will consider the solutions of
equation~(\ref{S_elliptic}) connected with the, so-called, minimal
surfaces. The minimal surface equation looks like as follows (see
e.g. Ref.~\cite{Osserman})
\begin{equation}
\label{minimal} \left(1 + S_{y}^{2} \right) S_{xx} + \left (1 +
S_{x}^{2} \right) S_{yy} - 2 S_{x} S_{y} S_{xy} = 0.
\end{equation}
We restrict ourselves to the class of solutions of
equation~(\ref{S_elliptic}) which are also harmonic on the
$xy-$plane, that is
\begin{equation}
\label{harmonic} S_{xx} + S_{yy} = 0.
\end{equation}
Using the condition~(\ref{harmonic}) in
equation~(\ref{S_elliptic}), one gets the equation
\begin{equation}
\label{S_elliptic_min} S_{x}^{2} S_{xx} + S_{y}^{2} S_{yy} + 2
S_{x} S_{y} S_{xy} = 0,
\end{equation}
for any elliptic intensity law. It is straightforward to check
that equations~(\ref{S_elliptic_min}) and~(\ref{minimal}) coincide
for harmonic solutions. In other words, a class of solutions of
equation~(\ref{S_elliptic}), whatever is the inverse intensity law
$I_{0} = I_{0}(u)$, is just given by the class of the harmonic
minimal surfaces. An important result for us is that the only
non-trivial harmonic minimal surface in Cartesian coordinates is
the helicoid~\cite{Osserman}. It can be written as follows
\begin{equation}
\label{helicoid} S = K \arctan \left (\frac{\beta \tau + y}{x}
\right),
\end{equation}
where $K$ is an arbitrary constant. One can check that the
function $S$ given by~(\ref{helicoid}) satisfies
equations~(\ref{minimal}), (\ref{harmonic})
and~(\ref{S_elliptic_min}) simultaneously. Equation of
corresponding wavefronts is
\begin{equation}
\label{helicoid_wavefronts} \tilde{S} \equiv \omega^{\alpha} \tau
+ K \arctan \left (\frac{\beta \tau + y}{x} \right) =
\textup{const},
\end{equation}
where the constant $K$ is the ``pitch'' of the helicoid. In
particular, the expression~(\ref{helicoid_wavefronts}) describes
the edge-screw dislocations discussed first time experimentally by
Brynghdal in 1973~\cite{Bryngdahl} and theoretically by Nye and
Berry~\cite{Berry} in 1974. We highlight that helicoidal
wavefronts exist in both linear and (nonlocal) nonlinear regimes.
They are associated with singularity of the phase ({\em phase
defects}) which appears as topological defects of the
interferograms. These class of phase defects has important
phenomenological consequences connected to the {\em optical
vortices}~\cite{Vaughan,Baranova,Coullet,Swartzlander} (see also
Refs.~\cite{Soskin1,Soskin2} and references therein). If we
assume, for simplicity, $\beta =0$ in~(\ref{helicoid}), one has
the pure screw dislocations. The complex gradient associated with
the helicoid
\begin{equation}
w = - i \frac{K}{z}
\end{equation}
is a meromorphic function on $z$ with the simple pole singularity
at the origin. Nevertheless, normal vector to the wavefront, whose
components coincides (up to a sign) with real and imaginary parts
of $w$, is not defined. Indeed, vector $w = (S_{x}, -S_{y})$ has
no limit as $\left (x, y\right) \to 0$. Figure~\ref{figure2} shows
examples of helicoidal wavefronts~(\ref{helicoid_wavefronts})
usually parametrized as follows
\begin{equation}
x = v \cos t, \quad{}y = v \sin t, \quad{}z = K t.
\end{equation}
One-start helicoid shown in figure~\ref{figure2}a is obtained for
$0< v < + \infty$; two-start helicoids shown in
figure~\ref{figure2}b,c are obtained for $- \infty < v < +
\infty$. Refractive index
\begin{equation}
\label{helicoidal_u} u = \frac{K^{2}}{4 (x^{2}+ y^{2})}
\end{equation}
has cylindrical symmetry around $\tau-$axis and displays
divergence at $z = 0$. This means that around the origin geometric
optics approximation fails and wave effects become relevant. In
particular, necessary condition for the existence of singular
wavefronts is that intensity vanishes where phase function is
singular~\cite{Berry}. Indeed, in this region the interference
phenomenon is no more negligible and can realize this condition.

It is interesting to evaluate the effect of the present class of
nonlocal perturbations
\[
S_{\tau} = \varphi \left(S_{x},S_{y},x,y,\tau \right)
\]
discussed in the sections~\ref{sec_generalmodel}
and~\ref{sec_paraxial}, on the pure ($\beta = 0$) helicoidal
wavefront~(\ref{helicoid}). In particular we note that the
solution~(\ref{helicoid}) of the equation~(\ref{S_elliptic_min})
is defined up to an additive arbitrary function of $\tau-$variable
$\psi(\tau)$, such that $\psi' = \varphi(\tau)$. Then, one has
\begin{equation*}
\label{S_helicoid} S = K \arctan \left(y/x \right) + \psi(\tau).
\end{equation*}
Let us observe that the function $\psi(\tau)$ is specified by the
function $\varphi$ evaluated on the helicoid. For example, in the
case of the first degree nonlocality one has (see
equations~(\ref{firstnonlocal_phi_hp})
and~(\ref{firstnonlocal_u_hp}))
\begin{gather}
\label{firstnonlocal_helicoid}
\begin{aligned}
\psi'(\tau) &= \alpha_{1} S_{x} + \alpha_{2} S_{y} \\
u_{\tau} &= \left(\alpha_{1} u \right)_{x} + \left(\alpha_{2}u
\right).
\end{aligned}
\end{gather}
Evaluating equations~(\ref{firstnonlocal_helicoid}) on the
helicoid one gets
\begin{gather}
\label{helicoidal_nonlocal_data}
\begin{aligned}
\left (x^{2} + y^{2} \right) \alpha_{1x} &= x \alpha_{1} + y
\alpha_{2} \\
- y \alpha_{1} + x \alpha_{2} &= \varphi(\tau) \left(x^{2} + y^{2}
\right).
\end{aligned}
\end{gather}
First degree nonlocal deformations of the helicoid exist only for
the nonlocal data $\alpha_{1}$ and $\alpha_{2}$ satisfying the
system~(\ref{helicoidal_nonlocal_data}). For example, trivial
solutions $\alpha_{1} = \alpha_{2} = \varphi = 0$ corresponds to
the local case discussed above. A simple non-trivial solution
$\alpha_{1} = - \gamma y$, $\alpha_{2} = \gamma x$ and $\psi' =
\gamma$, where $\gamma$ is an arbitrary constant, provides us with
the following wavefront
\begin{equation}
\label{helicoid_stretch} \tilde{S} \equiv \omega^{\alpha} \tau +
\frac{K}{1 + \omega^{-2 \nu} \gamma} \arctan \left(\frac{y}{x}
\right) = \textup{const}.
\end{equation}
For $\gamma = 0$ equation~(\ref{helicoid_stretch}) coincides with
equation~(\ref{helicoid_wavefronts}). As a consequence of nonlocal
response the helicoid's pitch is compressed if $\gamma > 0$ and
stretched if $\gamma < 0$.

\section{Nonlocal perturbations and dVN hierarchy.}
\label{sec_dVN_perturbation} In this section we will discuss a
particular class of nonlocal perturbations which are connected
with an integrable hierarchy of PDEs, namely, the dispersionless
Veselov-Novikov hierarchy. The methods to solve these equations,
in particular, the reduction method based on the symmetry
constraints and the quasiclassical $\bar{\partial}-$dressing
approach will be considered in sections 6 and 7.

Using the complex variables $z = x+iy$, $\bar{z} = x-iy$, one
rewrites equations~(\ref{Phase_generalsystem}) as follows
\begin{align}
\label{c_geom_eikonal_complex}
&S_{z} S_{\bar{z}} = u(z,\bar{z},\tau) \\
\label{c_geom_z_direction_complex} & S_{\tau} = \varphi\left
(S_{z},S_{\bar{z}},z,\bar{z},\tau \right).
\end{align}
The compatibility condition of
equations~(\ref{Phase_generalsystem}) imposes constraints on the
possible forms of the function $\varphi$, namely
\begin{equation}
\label{c_geom_compatibility} S_{\bar{z}} \varphi_{z} + S_{z}
\varphi_{\bar{z}} + u_{z} \varphi' + u_{\bar{z}} \varphi'' =
u_{\tau},
\end{equation}
where
\begin{equation}
\varphi' = \der{\varphi}{S_{z}}\left(z,\bar{z};S_{z},S_{\bar{z}}
\right),~~~~~~\varphi'' =
\der{\varphi}{S_{\bar{z}}}\left(z,\bar{z};S_{z},S_{\bar{z}}
\right).
\end{equation}
The simplest cases have been already discussed above and
corresponds to the local and first degree nonlocal responses,
given by the function $\varphi$ of the forms~(\ref{local_hp})
and~(\ref{firstnonlocal_phi_hp}) respectively. The quadratic case
\begin{equation}
\label{c_geom_quadratic_case} \varphi = \alpha S_{z}^{2} +
\bar{\alpha} S_{\bar{z}}^{2}+\beta S_{z}+ \bar{\beta} S_{\bar{z}}+
\gamma + \bar{\gamma},
\end{equation}
is equivalent to the linear one. Indeed, the consistency of
equations~(\ref{c_geom_compatibility})
and~(\ref{c_geom_eikonal_complex}) implies
\[
\alpha = 0,\quad{} \beta = \beta(z), \quad{} \gamma = const,
\]
and the~(\ref{c_geom_quadratic_case}) is reduced to
the~(\ref{firstnonlocal_phi_hp}).

The cubic function
\begin{equation}
\label{c_geom_cubic_case} \varphi = \alpha S_{z}^{3} +
\bar{\alpha} S_{\bar{z}}^{3}+ \beta S_{z}^{2}+\bar{\beta}
S_{\bar{z}}^{2}+ \gamma S_{z} + \bar{\gamma} S_{\bar{z}} + \delta
+ \bar{\delta}
\end{equation}
obeys equations~(\ref{c_geom_compatibility})
and~(\ref{c_geom_eikonal_complex}) if
\begin{eqnarray}
&&\alpha = \alpha(z),~~~~~~\beta = 0, \nonumber \\
&&\gamma_{\bar{z}} = - \alpha_{z} u - 3 \alpha u_{z},
\nonumber \\
&&\delta = const,
\end{eqnarray}
and one has the equation
\begin{equation}
\label{c_geom_u_cubic} u_{\tau} = \left(\gamma u\right)_{z} +
\left(\bar{\gamma} u \right)_{\bar{z}}.
\end{equation}
In the particular case $\alpha=1$ and, consequently
$\gamma_{\bar{z}} = -3 u_{z}$, equation~(\ref{c_geom_u_cubic}) is
the dispersionless Veselov-Novikov (dVN) equation introduced
in~\cite{Krichever,Konopelchenko4}.

In a similar way, repeating the procedures for higher degree
polynomials, one can construct higher order nonlocal
perturbations.

Besides, if we formally admit all possible degrees of $S_{z}$ and
$S_{\bar{z}}$ in the right hand side of
equation~(\ref{c_geom_z_direction_complex}), one has an infinite
family of nonlinear equations, governing the nonlocal deformations
of the wavefronts and ``refractive index'' $u$.

Moreover, the polynomial dependence of these deformations should
be compatible with certain constraints. The request that
equation~(\ref{c_geom_z_direction_complex}) respects the symmetry
for phase inversion, $S \rightarrow - S$, of the eikonal
equation~(\ref{c_geom_eikonal_complex}) gives the hierarchy of
nonlocal deformations of the form
\begin{equation}
\sum_{m=1}^{n} u_{m} S_{z}^{2m-1} + \sum_{m=1}^{n} \bar{u}_{m}
S_{\bar{z}}^{2m-1}.
\end{equation}
Note that the constant terms which have appeared
in~(\ref{c_geom_u_cubic})is, in fact, irrelevant. Hence, these
polynomial deformations assume the general form
\begin{equation}
\label{c_geom_higherpolynomials} S_{\tau} = \sum_{m=1}^{n}
\left(u_{m} S_{z}^{2m-1} + \bar{u}_{m} S_{\bar{z}}^{2m-1}\right),
~~~~~n=1,2,3,\dots
\end{equation}
where $u_{m}$ are certain functions on $u$.

In the case $u_{n}=1$ one gets the dVN equation mentioned above
($n=2$) and the so-called dVN hierarchy of nonlinear PDEs. This
means that the dVN hierarchy is associated with a specific class
of ``integrable" nonlocal perturbations. In particular, there is a
one-to-one correspondence between nonlocal degrees and the
equations of the dVN hierarchy. It is straightforward to verify
that the helicoidal wavefronts are not preserved under dVN
hierarchy nonlocal perturbations.

\section{Integrable systems.}
\label{sec_integrable} The celebrated Korteweg-de-Vries (KdV)
equation
\begin{equation}
\label{kdv_ex} u_{t} + 6 u u_{x} + u_{xxx} = 0
\end{equation}
has been introduced to describe long, one-dimensional, small
amplitude shallow water wave propagation~\cite{Segur}. By a
numerical analysis Zabusky and Kruskal~\cite{Zabusky} observed
main feature of the solitons, that is they ``pass one another
without losing their identity" . Solitons, can be seen as a purely
nonlinear phenomenon, where the linear dispersion is ``balanced"
by the nonlinearity. Fundamental step has been taken in 1967 with
the famous paper by Gardner, Green, Kruskal and Miura, with the
introduction of the so-called inverse spectral transform method
(IST). The IST approach works as a nonlinear Fourier transform
according to the following scheme
\begin{align*}
u(x,0) &\to \textup{{\tt Scattering data at} $t=0$} \to
\textup{\small{{\tt evolution}}} \\&\to \textup{{\tt Scattering
data at} $t=t'$} \to  u(x,t').
\end{align*}
By means of the spectral transform the initial profile $u(x,0)$ is
associated with a set of spectral data at the time $t=0$. In the
space of spectral data the system results to be linearized and the
evolution is trivial. Then, once the spectral data at the time
$t=t'$ are obtained, inverse spectral transform gives the evolved
profile $u(x,t')$, solution of nonlinear PDE. The method is based
on the representation of a nonlinear PDE as the compatibility
condition of a pair of linear problems. In the KdV case the linear
system is
\begin{subequations}
\label{lax_kdv}
\begin{align}
\label{lax_kdv1} &-\dersec{\psi}{x} - u(x,t) \psi  = \lambda^{2} \psi, \\
\label{lax_kdv2} &\der{\psi}{t} + \derthree{\psi}{x} - 3
\left(\lambda^{2} - u \right) \der{\psi}{x} - 4 i \lambda^{3} \psi
= 0
\end{align}
\end{subequations}
where $\lambda$ is usually called spectral parameter. The
system~(\ref{lax_kdv}) can be regarded also as the compatibility
condition of the couple of operators ({\em Lax
pair})~(\ref{lax_kdv})
\begin{equation}
\left [L_{1},L_{2} \right] = 0
\end{equation}
where
\begin{align}
 L_{1} &=  \dersec{}{x} + u(x,t) + \lambda^{2} \\
 L_{2} &= \der{}{t} + \derthree{}{x} - 3 \left(\lambda^{2} - u
\right) \der{}{x} - 4 i \lambda^{3}
\end{align}
The nonlinear Schr\"odinger (NLS) equation
\begin{equation}
\label{NLS} i q_{t} + q_{xx} + 2 s \left|q \right|^{2} q = 0
\end{equation}
and sine-Gordon (SG) equation
\begin{equation}
\label{SG} u_{tt}- u_{xx} + \sin u = 0
\end{equation}
are other remarkable examples of $(1+1)D$ nonlinear integrable
PDEs. NLS equation has applications, for instance, in nonlinear
optics~\cite{Boyd} and Bose-Einstein condensation~\cite{Bosecond}
and SG equation appears in geometry to describe negative curvature
surfaces~\cite{Eisenhart}, in physics in the study of flux
propagation in Josephson junctions, nonlinear optics and quantum
field theory (see e.g.~\cite{Segur}). Thereafter, the method has
been generalized to $(2+1)-$dimensional soliton equations, such as
the Kadomtsev-Petviashvili equation
\begin{gather}
\label{kp_ex}
\begin{aligned}
u_{t} + u_{xxx} + \frac{3}{2} u u_{x} + \frac{3}{4}  v_{y} &= 0 \\
v_{x} &= u_{y}.
\end{aligned}
\end{gather}
Linear problems for~(\ref{kp_ex}) are
\begin{subequations}
\label{lax_kp}
\begin{align}
\psi_{y} &= \psi_{xx} + u \psi \\
\psi_{t} &= \psi_{xxx} + \frac{3}{2} u \psi_{x} +
\left(\frac{3}{2} u_{x} + \frac{3}{4} v \right) \psi = 0
\end{align}
\end{subequations}
where $v_{x} = u_{y}$. Equation~(\ref{kp_ex}) is the $(2+1)D$
generalization of KdV describing weakly two-dimensional long,
shallow water waves. Another ($2+1$)D integrable generalization of
KdV equation is given by the Nizhnik-Veselov-Novikov (NVN)
equation
\begin{equation}
\label{dvn_ex} u_{t} + c_{1} u_{\xi \xi \xi} + c_{2} u_{\eta \eta
\eta} - 3 c_{1} \left(u \partial_{\xi}^{-1} u_{\eta}
\right)_{\eta} - 3 c_{2} \left(u \partial_{\eta}^{-1}u_{\xi}
\right)_{\xi} = 0
\end{equation}
where $\xi = x + s y$, $\eta = x - s y$. Equation~(\ref{dvn_ex})
has been introduced by Nizhnik in the case $s = 1$~\cite{Nizhink}
and by Veselov and Novikov in the case $s = i$,
$c_{1}=c_{2}=1$~\cite{VNoriginal}. Unlike the KP equation no
physical applications of the NVN equation is known at the moment.
The Veselov-Novikov (VN) equation ($z = x + i y$)
\begin{subequations}
\begin{align}
\label{VN_full} u_{t} &= \left(u V \right)_{z} + \left(u \bar{V}
\right)_{\bar{z}}
+ u_{zzz} + u_{\bar{z}\bar{z}\bar{z}} \\
V_{\bar{z}}& = -3 u_{z},
\end{align}
\end{subequations}
is equivalent to the compatibility condition for equations
\begin{align*}
&\psi_{z\bar{z}} = u \psi \\
&\psi_{t} = \psi_{zzz} + \psi_{\bar{z}\bar{z}\bar{z}} +
\left(V\psi_{z} \right) + \left(\bar{V}\psi_{\bar{z}} \right).
\end{align*}
Note also that the local $(2+1)D$ NLS equation is not integrable
by the IST method.

Recently the, so-called, quasi-classical limit (or dispersionless)
limit of soliton equations has attracted a great interest (see
e.g.~\cite{Kodama1})-\cite{Interface_2002}. Dispersionless limit
of soliton equations is performed formally introducing a slow
variable expansion by the substitution
\begin{equation}
t_{n} \to t_{n}/\epsilon
\end{equation}
and introducing the asymptotic expansion of the form
\begin{equation}
u \left(\frac{t_{n}}{\epsilon} \right) = u \left(t_{n} \right) + O
\left(\epsilon \right), \quad{} \epsilon \to 0.
\end{equation}
where $t_{n}$ the set of the ``time" variables. For example, in
the KdV equation case one identify $t_{1} =x$, $t_{2} = y$, while
for KP equation it is $t_{1} =x$, $t_{2}=y$ and $t_{3} = t$.

In terms of the linear problems, this procedure corresponds to the
following representation of the function $\psi$
\begin{equation*}
\psi = \psi_{0} \; e^{\frac{S}{\epsilon}}
\end{equation*}
where
\begin{equation*}
S \left(\lambda; \frac{t_{n}}{\epsilon} \right) = S \left(t_{n}
\right) + O \left(\epsilon \right).
\end{equation*}
For example, in the leading order equation~(\ref{kp_ex}) is the
dispersionless KP (dKP) equation
\begin{gather}
\label{dKP_ex}
\begin{aligned}
u_{t} &= \frac{3}{2} u u_{x} + \frac{3}{4} v_{y} \\
v_{x} &= u_{x}
\end{aligned}
\end{gather}
and the linear problems~(\ref{lax_kp}) becomes the Hamilton-Jacobi
equations
\begin{gather}
\label{dKP_Lax}
\begin{aligned}
S_{y} &= S_{x}^{2} + u \\
S_{t} &= S_{x}^{3} + \frac{3}{2} u S_{x} + \frac{3}{2} u_{x} +
\frac{3}{4} v.
\end{aligned}
\end{gather}
The dKP equation is known in physics as Zabolotskaya-Khokhlov
equation~\cite{Khokhlov}.

In the following, in connection with the model of nonlocal
nonlinear optics discussed above, we will focus on the
dispersionless Veselov-Novikov (dVN) equation. It looks like as
follows
\begin{gather}
\begin{aligned}
u_{t} &= \left(u V \right)_{z} + \left(u \bar{V} \right)_{\bar{z}}
\\
V_{\bar{z}} &= - 3 u_{z}.
\end{aligned}
\end{gather}
It is obtained as the compatibility condition of the following
system
\begin{gather}
\label{HJ_dVN}
\begin{aligned}
&S_{z} S_{\bar{z}} = u \\
&S_{t} = S_{z}^{3} + S_{\bar{z}}^{3} + V S_{z} + \bar{V}
S_{\bar{z}}.
\end{aligned}
\end{gather}
Dispersionless systems are of interest for both physical and
mathematical reasons. They have applications in hydrodynamincs,
magneto-hydrodynamics, Laplacian growth~\cite{Interface_PRL}. They
arise also in the framework of the Whitham averaging method for
calculation of small amplitude modulations of soliton equations
solutions~\cite{whithambook,Dubrovin}. Moreover, we will see in
the following that the dVN equation and the hierarchy associated
are relevant in the description of specific high frequency regimes
in nonlocal nonlinear optics.

Different approaches can be used to study dispersionless systems.
Here we will consider the reduction method based on the symmetry
constraints and the $\bar{\partial}-$dressing method. In
particular, their applications to the eikonal equation and the
nonlocal perturbations associated with the dVN equation will be
also discussed.

\section{Symmetry constraints.}
\label{sec_sym_contraints} Here we will recall shortly the
definition of symmetry constraint and discuss its specific
application to the dVN equation as an approach for construction of
($1+1$D) hydrodynamic type reductions. Let us consider a partial
differential equation for the scalar function $u = u(t) = u
\left(t_{1},t_{2},\dots \right)$
\begin{equation}
\label{general} F\left(u,u_{t_{i}},u_{t_{i}t_{j}},\dots \right) =
0,
\end{equation}
where $u_{t_{i}} = \partial u/\partial t_{i}$. By definition, a
symmetry of equation~(\ref{general}) is a transformation $u(t)
\rightarrow u'(t')$, such that $u'(t')$ is again a solution
of~(\ref{general}) (for more details see
\textit{e.g.},~\cite{Olver}). Infinitesimal continuous symmetry
transformations
\begin{equation}
t_{i}'=t_{i}+\delta t_{i};~~~~~u'=u + \delta u = u + \epsilon
u_{\epsilon}
\end{equation}
are defined by the linearized equation~(\ref{general})
\begin{equation}
\label{symmetry_def} L\delta u = 0,
\end{equation}
where $L$ is the Gateaux derivative of $F$
\begin{equation}
\label{gateaux} L\delta u := \left . \dertot{F}{\epsilon} \left(u
+ \epsilon u_{\epsilon},\der{}{t_{i}}\left(u+\epsilon
u_{\epsilon}\right),\dots),\dots \right) \right |_{\epsilon = 0}.
\end{equation}
Any linear superposition $\delta u =\sum_{k} c_{k} \delta_{k} u$
of infinitesimal symmetries $\delta_{k}u$ is, obviously, an
infinitesimal symmetry. By definition, a {\em symmetry constraint}
is a requirement that certain superposition of infinitesimal
symmetries vanishes, i.e.
\begin{equation}
\label{constraint_gen} \sum_{k} c_{k} \delta_{k} u = 0.
\end{equation}
Since null function is a symmetry of equation~(\ref{general}), the
constraint~(\ref{constraint_gen}) is compatible with
equation~(\ref{general}). Symmetry constraints allow us to select
a class of solutions which possess some invariance properties. For
instance, well-known symmetry constraint $\delta u = \epsilon
u_{t_{k}} = 0$, selects solutions which are stationary with
respect to the ``time''~$t_{k}$.

\subsection{The dKP equation.}
\label{sec_dKP} The linearized equation associate with the dKP
equation~(\ref{dKP_ex}) is
\begin{align}
 \label{symmetry_dKP}
(\delta u)_{t}& = \frac{3}{2} \left(u_{x} \delta u + u (\delta
u)_{x}\right) + \frac{3}{4} (\delta
\omega)_{y}, \nonumber\\
(\delta \omega)_{x}& = (\delta u)_{y}.
\end{align}
Solutions are infinitesimal symmetries of dKP.

\begin{theorem}
Suppose $S_{i}$ and $\tilde{S}_{i}$ are arbitrary solutions  of
the
  Hamilton-Jacobi equations~\eqref{HJ_dVN}. Then the quantity
  $$\delta u =
  \sum_{i=1}^{N} c_{i}\left(S_{i}-\tilde{S}_{i}\right)_{xx},$$
  where $c_{i}$
are arbitrary constants, is a symmetry of dKP equation.
\end{theorem}

{\proof It is straightforward to check that
${\left(S_{i}-\tilde{S}_{i}\right)}_{xx}$ satisfies
equation~(\ref{symmetry_dKP}). $\Box$ }

This type of symmetries has been introduced for the first time
in~\cite{Bogdanov}, within the quasiclassical
$\bar{\partial}$-dressing approach. As a simple example, let us
consider the following symmetry constraint
\begin{equation}
u_{x} = S_{xx}.
\end{equation}
Under this constraint the Hamilton-Jacobi system~(\ref{dKP_Lax})
gives rise to the following hydrodynamic type system (the
dispersionless nonlinear Schr\"odinger equation,
see~\cite{Zakharov}):
\begin{align}
\tilde{u}_{y}& = \left(\tilde{u}^{2} + u \right)_{x},  \nonumber\\
u_{y}& = 2 \left(\tilde{u} u \right)_{x}
\end{align}
where $\tilde{u} = \partial{S_{x}}/\partial{\lambda}$.

\subsection{Real dVN equation}
\label{sec_real_dVN}
Let us focus on the case of real-valued $u$. Infinitesimal
symmetries $\delta u$ of the dVN equation obey the equations
\begin{align}
\left(\delta u \right)_{t}& = \left(V \delta u + u \delta V
\right)_{z} + \left(\bar{V} \delta u + \delta \bar{V} u
\right)_{\bar{z}} \nonumber \\
V_{\bar{z}}& = -3 u_{z}; \quad{} \left(\delta V \right)_{\bar{z}}
= -3 \left(\delta u \right)_{z}.\label{linearized_dVN}
\end{align}

\begin{theorem}
Suppose $S_{i}$ and $\tilde{S}_{i}$ are  solutions  of the
  Hamilton-Jacobi
  equations~\textup{(\ref{HJ_dVN}).}
  Then the   quantity
\begin{equation}
\label{constraint_S} \delta u =
  \sum_{i=1}^{N} c_{i}\left(S_{i}-\tilde{S}_{i}\right)_{z\bar{z}},
\end{equation}
where $c_{i}$ are arbitrary constants, is a symmetry of the dVN
equation.
\end{theorem}

{\proof
 It is straightforward to check that
${\left(S_{i}-\tilde{S}_{i}\right)}_{z\bar{z}}$ satisfies
equation~(\ref{linearized_dVN}). $\Box$}

In particular, one can choose $S_{i} = S(\lambda = \lambda_{i})$
and $\tilde{S}_{i} = S(\lambda = \lambda_{i} + \mu_{i})$. In the
case $\mu_{i} \to 0$ and $c_{i} = {\tilde{c}_{i}}/{\mu_{i}}$, one
has the class of symmetries given by
\begin{subequations}\label{constraint_Phi}
\begin{align}
\delta u& = \sum_{i=1}^{N} \tilde{c}_{i} \phi_{iz\bar{z}} \\
\phi_{i}& = \der{S}{\lambda}(\lambda = \lambda_{i}).
\end{align}
\end{subequations}
In what follows we will discuss three particular cases of real
reductions, providing real solutions of the dVN equation.

If $S$ is a solution of Hamilton-Jacobi equations~(\ref{HJ_dVN}),
then $-\bar{S}$ is a solution as well. Thus, for real-valued $S$
($S= \bar{S}$), specializing constraint~(\ref{constraint_S}) for
$N=1$, we have a simple constraint
\begin{align}
\label{case1} u_{x} = \left(S \right)_{z\bar{z}}.
\end{align}
Let us introduce the functions $\rho_{1}:=S_{x}$ and
$\rho_{2}:=S_{y}$. Thus, the symmetry constraint~(\ref{case1}) can
be written as follows
\begin{equation}
\label{case_I} u_{x} =\frac{1}{4} \left(S_{xx}+S_{yy} \right) =
\frac{1}{4}\left( \rho_{1x} + \rho_{2y} \right).
\end{equation}
In order to analyze constraint~(\ref{case_I}) it is more
convenient to consider equations~(\ref{HJ_dVN}) in Cartesian
coordinates $(x$, $y)$, \textit{i.e.},
\begin{subequations}
\begin{align}
\label{dVN_Lax_cartesian1}
&S_{x}^{2} + S_{y}^{2} = 4 u \\
\label{dVN_Lax_cartesian2} &S_{t} = \frac{1}{4} S_{x}^{3}
-\frac{3}{4} S_{x}S_{y}^{2} + V_{1} S_{x} + V_{2} S_{y},
\end{align}
\end{subequations}
where $V = V_{1} + i V_{2}$, while the dVN equation acquires the
form
\begin{subequations}
\label{dVN_cartesian}
\begin{align}
\label{dVN_cartesian1} &u_{t} = \left(u V_{1} \right)_{x} +
\left(u V_{2} \right)_{y}\\
\label{dVN_cartesian2}
&V_{1x} - V_{2y} = - 3 u_{x} \\
\label{dVN_cartesian3} &V_{2x} + V_{1y} = 3 u_{y}.
\end{align}
\end{subequations}
Substituting~(\ref{dVN_Lax_cartesian1}) in~(\ref{case_I}), one
obtains the following hydrodynamic type system
\begin{equation}
\label{hydro_system1} \left( \begin{array}{cc}
\rho_{1}  \\ \rho_{2} \\
\end{array} \right)_{y} =
\left( \begin{array}{cccc}
0 & 1 \\
2 \rho_{1} - 1 & 2 \rho_{2}\\
\end{array} \right)
\left( \begin{array}{cc}
\rho_{1}  \\ \rho_{2} \\
\end{array} \right)_{x}.
\end{equation}

Now, let us focus on definition $V_{\bar z} := - 3 u_{z}$.
Differentiating it with respect to $x$, using
constraint~(\ref{case1}) and equations~(\ref{hydro_system1}), one
obtains the equations
\begin{align*}
V_{1x}& = - \frac{3}{2} \rho_{1x} + \frac{3}{4} \left(
\rho_{1}^{2} +
  \rho_{2}^{2}\right)_{x}  \\
V_{2x}& = \frac{3}{2} \rho_{2x},
\end{align*}
which can be trivially integrated providing the following explicit
formulas for $V_{1}$ and $V_{2}$ in terms of $\rho_{1}$ and
$\rho_{2}$:
\begin{align}
V_{1}& = - \frac{3}{2} \rho_{1} + \frac{3}{4} \left( \rho_{1}^{2}
+
  \rho_{2}^{2}\right) \nonumber  \\
V_{2}& = \frac{3}{2} \rho_{2}.\label{case1_V}
\end{align}
At this point we can derive $t$-dependent equations for $\rho_{1}$
and
 $\rho_{2}$.
Differentiating equation~(\ref{dVN_Lax_cartesian2}) and using
~(\ref{hydro_system1})~and~(\ref{case1_V}), one obtains the system
\begin{equation}
\label{hydro_system2}
\left( \begin{array}{cc} \rho_{1}  \\ \rho_{2} \\
\end{array} \right)_{t} =\left( \begin{array}{cc}
A_{11} & A_{12}  \\ A_{21} & A_{22} \\
\end{array} \right)\left( \begin{array}{cc} \rho_{1}  \\ \rho_{2} \\
\end{array} \right)_{x},
\end{equation}
where
\begin{eqnarray*}
&A_{11} = 3 \rho_{1} \left( \rho_{1} -1 \right) ,  &A_{12} = 3 \rho_{2}, \\
&A_{21} = 3 \rho_{2} \left (2 \rho_{1}-1 \right),  &A_{22} = 3
  \rho_{1} \left(\rho_{1}-1 \right) + 6 \rho_{2}^{2}.
\end{eqnarray*}
Common solutions $\left (\rho_{1},\rho_{2} \right)$ of the
systems~(\ref{hydro_system1}) and~(\ref{hydro_system2}) provides
us with the solution $u = \left( \rho_{1}^{2}+ \rho_{2}^{2}
\right)/4$ of the dVN equation~(\ref{dVN_cartesian}).

\subsection{Admissible intensity laws.}
\label{sec_admissible} The analysis of the compatibility between
the class of nonlocal perturbations associated with the dVN
hierarchy and the intensity conservation
law~(\ref{transverse_system_conservation}) shows that there are no
nontrivial solutions for an arbitrary form of the intensity law.
In particular, intensity law appears to be a quite restrictive
constraint making the problem of compatible nonlocal responses
rather non-trivial. Nevertheless, it is possible to see, by an
explicit example, that there exists nontrivial intensity laws such
that the system~(\ref{phase}) and the intensity conservation
law~(\ref{transverse_system_conservation}) are compatible with the
dVN hierarchy. Here, as illustrative example, we focus on the
third degree of nonlocality associated with the dVN equation.

In order to do that, we consider hydrodynamic type reductions of
dVN equation. They have been found using symmetry constraint of
the form~\cite{Bogdanov}
\begin{equation*}
\nabla_{\bot}^{2}S = u_{x}
\end{equation*}
or equivalently $u_{x} = S_{z \bar{z}}$. Under such a constraint
one gets the hydrodynamic type
system~(\ref{hydro_system1}),~(\ref{hydro_system2}).

Looking for solutions of the system~(\ref{hydro_system1})
and~(\ref{hydro_system2}) such that $p_{2} = p_{2} (p_{1})$, one
finds that $p_{1}$ and $p_{2}$ are given implicitly by the
following algebraic system
\begin{align}
\label{dVN_solutions} &x + G' y + H' z - \Phi \left (p_{1} \right
)
= 0 \\
\label{p2_algebraic} &p_{2} = \frac{1}{2} \left [ q + \frac{2 c -
\log \left (q + \sqrt{1 + q^{2}} \right )}
{\sqrt{1 + q^{2}}} \right ]  \\
&q = p_{2} \pm \sqrt{p_{2}^{2} + 2 p_{1} -1} \nonumber
\end{align}
where $c$ is an arbitrary constant, $G = p_{2}(p_{1})$, $H =
p_{1}^{3} - \frac{3}{2} p_{1}^{2} + \frac{3}{2} p_{2}^{2}(p_{1})$
and `prime' means the derivative with respect to $p_{1}$.

Differentiating eikonal equation~(\ref{eikonal}) with respect to
$x$ and taking into account that $u_{x} = \nabla_{\bot}^{2}S$, one
gets
\begin{equation}
\label{eikonal_constraint} \nabla_{\bot}\left (- \frac{S_{x}}{2}
\right) \cdot \nabla_{\bot}S + \nabla_{\bot}^{2}S = 0.
\end{equation}
Comparing equation~(\ref{eikonal_constraint}) with intensity
conservation equation~(\ref{transverse_system_conservation}),
which can be written equivalently as follows
\begin{equation}
\label{intensity_conservation} \nabla_{\bot}S \cdot
\nabla_{\bot}\left(\log I_{0} \right) + \nabla_{\bot}^{2}S = 0
\end{equation}
 one gets the following simple
relation among intensity and $p_{1}-$component of the gradient
\begin{equation}
\label{intensity_symmetry} I_{0} = C\; e^{- \frac{p_{1}}{2}},
\end{equation}
where $C$ is an arbitrary real constant. Finally, eikonal equation
provides us with intensity law
\begin{equation}
\label{refractive_symmetry} u (I_{0}) = \left (\log
\frac{I_{0}}{C} \right)^{2} + \frac{1}{4} \left [ p_{2} \left (- 2
\log \frac{I_{0}}{C} \right) \right ]^{2},
\end{equation}
where last term in r.h.s. is given by the algebraic
relation~(\ref{p2_algebraic}).

\section{The quasiclassical $\bar{\partial}$-dressing method.}
\label{sec_QC_Dbar}

\subsection{The eikonal equation.}
\label{sec_QC_eikonal} In this and next sections we will
demonstrate that the plane eikonal equation and dVN hierarchy both
for complex and real refractive indices are treatable by the
quasiclassical $\bar{\partial}$-dressing method.

The quasiclassical $\bar{\partial}$-dressing method is based on
the nonlinear Beltrami equation \cite{ragnisco,Konopelchenko3}
\begin{equation}
\label{Beltrami_nonlinear} S_{\bar{\lambda}} =
W\left(\lambda,\bar{\lambda};S_{\lambda} \right)
\end{equation}
where $S(\lambda,\bar{\lambda})$ is a complex valued function,
$\lambda$ is the complex variable $S_{\lambda} = \partial
S/\partial \lambda$ and $W$ (the quasiclassical
$\bar{\partial}$-data)
 is an analytic function of $S_{\lambda}$
\begin{equation}
\label{kernel} W\left(\lambda,\bar{\lambda},S_{\lambda}\right) =
\sum_{n=0}^{\infty}w_{n}(\lambda,\bar{\lambda})
\left(S_{\lambda}\right)^{n},
\end{equation}
with some, in general, arbitrary functions
$w_{n}(\lambda,\bar{\lambda})$.

To construct integrable equations one has to specify the domain
$G$ (in the complex plane $\mathbb{C}$) of support for the
function $W$ ($W=0$, $\lambda \in \mathbb{C}/G$) and look for
solution of~(\ref{Beltrami_nonlinear}) in the form $S = S_{0} +
\tilde{S}$, where the function $S_{0}$ is analytic inside $G$,
while $\tilde{S}$ is analytic outside $G$. In order to construct
the eikonal equation on the plane~\cite{Moro2}, we choose $G$ as
the ring ${\cal D} = \left \{\lambda \in \mathbb{C} :~ \frac{1}{a}
< |\lambda| < a \right\}$,  where $a$ is an arbitrary real number
($a
> 1$), and select $B-type$ solutions satisfying the constraint
\begin{equation}
\label{B_symmetry} S\left(- \lambda,-\bar{\lambda}\right)= -
S\left(\lambda,\bar{\lambda}\right).
\end{equation}
Then, we choose
\begin{equation}
\label{S_0} S_{0} = z \lambda + \frac{\bar{z}}{\lambda}.
\end{equation}
Due to the analyticity of $\tilde{S}$ outside the ring and the
property~(\ref{B_symmetry}) one has
\begin{equation}
\tilde{S} = \sum_{n=0}^{\infty}
\frac{S_{2n+1}^{(\infty)}}{\lambda^{2n+1}};~~~~~\lambda
\rightarrow \infty,
\end{equation}
and
\begin{equation}
\tilde{S} = \sum_{n=0}^{\infty}S_{2n+1}^{(0)}
\lambda^{2n+1}~~~~~\lambda \rightarrow 0.
\end{equation}
In particular, $\tilde{S}\left(0,0 \right) = 0$. \\
An important property of the nonlinear
$\bar{\partial}$-problem~(\ref{Beltrami_nonlinear}) is that the
derivatives $f = S_{t}$ of $S$ with respect to any independent
variable $t$, obeys the linear Beltrami equation
\begin{equation}
\label{Beltrami} f_{\bar{\lambda}}
=W'\left(\lambda,\bar{\lambda};S_{\lambda}\right) f_{\lambda}
\end{equation}
where $W'\left (\lambda,\bar{\lambda};\phi \right) =
\der{W}{\phi}\left(\lambda,\bar{\lambda};\phi \right)$.
Equations~(\ref{Beltrami}) has two basic properties, namely, 1)
any differentiable function of solutions $f_{1},\dots,f_{n}$ is
again a solution; 2) under certain mild conditions on $W'$, a
bounded solution $f$ which is equal to zero at certain point
$\lambda_{0} \in \mathbb{C}$, vanishes identically (Vekua's
theorem) \cite{Vekua}.

These two properties allows us to construct an equation of the
form $\Omega\left(S_{z},S_{\bar{z}}\right) = 0$ with certain
function $\Omega$. Indeed, taking into account~(\ref{S_0}), one
has
\begin{eqnarray}
S_{z}  = \lambda + \tilde{S}_{z}, \\
S_{\bar{z}} = \frac{1}{\lambda} + \tilde{S}_{\bar{z}},
\end{eqnarray}
i.e. $S_{z}$ has a pole at $\lambda = \infty $, while
$S_{\bar{z}}$ has a pole at $\lambda = 0$. The product $S_{z}
S_{\bar{z}}$ is again a solution of the linear Beltrami
equation~(\ref{Beltrami}) and it is bounded on the complex plane
since
\begin{equation}
\label{bounded} S_{z}S_{\bar{z}} = 1+ \frac{1}{\lambda}
\tilde{S}_{z} + \lambda \tilde{S}_{\bar{z}} +
\tilde{S}_{z}\tilde{S}_{\bar{z}}
\end{equation}
where $\tilde{S}_{\bar{z}}\left(\lambda=0 \right) = 0$ and
$\lambda \tilde{S}_{\bar{z}} \rightarrow S_{1\bar{z}}^{(\infty)}$
as $\lambda \rightarrow \infty$.

Subtracting $1+S_{1\bar{z}}^{(\infty)}$ from the r.h.s of
equation~(\ref{bounded}), one gets a solution of
equation~(\ref{Beltrami}) which is bounded in $\mathbb{C}$ and
vanishes as $\lambda \rightarrow \infty$. According to the Vekua's
theorem it is equal to zero for all $\lambda$. Thus we get the
equation
\begin{equation}
\label{eikonal2} S_{z} S_{\bar{z}} = u(z,\bar{z})
\end{equation}
where
\begin{equation}
\label{residual} u = 1 + S_{1,\bar{z}}^{(\infty)}.
\end{equation}
In the Cartesian coordinates $x,y$ defined by $z=x+iy$,
equation~(\ref{eikonal2}) is the standard two-dimensional eikonal
equation discussed above
\begin{equation}
\label{eikonal3} \left(\nabla_{\bot} S \right)^{2} = 4 u
\end{equation}
where we recall that $\nabla_{\bot} = \left(\der{}{x}, \der{}{y}
\right)$.

Using the $\bar{\partial}$-problem~(\ref{Beltrami_nonlinear}), one
can, in principle,  construct solutions of
equation~(\ref{eikonal2}). So, the quasiclassical
$\bar{\partial}$-dressing method allows us to treat the plane
eikonal equation~(\ref{eikonal2}) in a way similar to dKP and
d2DTL equations. We note that the phase function $S$
in~(\ref{eikonal2}) depends also on the complex variables
$\lambda$ and $\bar{\lambda}$. Curves $S(z,\bar{z})=const$ define
wavefronts. The $\bar{\partial}$-dressing approach provides us
also with the equation of light rays. Indeed, since r.h.s.
of~(\ref{eikonal3}) does not depend on $\lambda$ and
$\bar{\lambda}$, the differentiation of~(\ref{eikonal3}) with
respect of $\lambda$ (or $\bar{\lambda}$) gives
\begin{equation}
\label{scalar} \nabla S \cdot \nabla \phi = 0
\end{equation}
where $\phi = S_{\lambda}$ (or $\phi = S_{\bar{\lambda}}$). So,
the curves $S=const$ and $\phi=const$ are reciprocally orthogonal
and, hence, the latter ones are nothing but the trajectories of
propagating light. Thus, the $\bar{\partial}$-dressing approach
provides us with all characteristics of the propagating light on
the plane. Note that any differentiable function $\phi
\left(S_{\lambda}, S_{\bar{\lambda}}\right)$ is the solution of
equation~(\ref{scalar}) too.

In general, within the $\bar{\partial}$-dressing approach one has
a complex-valued phase function $S$ and, consequently, a complex
refractive index. To guarantee the reality of $u$, it is
sufficient to impose the following constraint on $S$
\begin{equation}
\label{parity_circle}
\overline{S}\left(\lambda,\bar{\lambda}\right) =
S\left(\frac{1}{\bar{\lambda}},\frac{1}{\lambda} \right).
\end{equation}
Indeed, taking the complex conjugation of
equation~(\ref{eikonal2}), using the differential consequences
(with respect to $z$ and $\bar{z}$), of the above constraint and
taking into account the independence of the l.h.s. of
equation~(\ref{eikonal2}) on $\lambda$,$\bar{\lambda}$, one gets
\begin{equation}
\bar{u}\left (x_{n}\right) =
\bar{S}_{\bar{z}}\left(\lambda,\bar{\lambda} \right)
\bar{S}_{z}\left(\lambda,\bar{\lambda} \right) =
S_{\bar{z}}\left(\frac{1}{\bar{\lambda}},\frac{1}{\lambda} \right)
S_{z}\left(\frac{1}{\bar{\lambda}},\frac{1}{\lambda} \right) =
u\left(x_{n} \right),
\end{equation}
i.e. the ``refractive index'' $u$ is real one. The
constraint~(\ref{parity_circle}) leads to the relations
$\bar{S}_{2n+1}^{(0)} = S_{2n+1}^{(\infty)}$. Moreover, this
constraint implies also that the function $S$ is real-valued on
the unit circle $\left|\lambda \right| = 1$ ($\overline{S}
\left(\lambda,\bar{\lambda} \right) =
S\left(\lambda,\bar{\lambda}\right)$, $|\lambda| =1$). This
provides us with the physical wavefronts.

The $\bar{\partial}$-approach reveals also the connection between
geometrical optics and the theory of the, so-called,
quasiconformal mappings on the plane. Quasi-conformal mappings
represents themselves a very natural and important extension of
the well-known conformal mappings (see e.g. \cite{Ahlfors}). In
contrast to the conformal mappings the quasi-conformal mappings
are given by non-analytic functions, in particular, by solutions
of the Beltrami equation.

A solution of the nonlinear Beltrami
equation~(\ref{Beltrami_nonlinear}) defines a quasi-conformal
mapping of the complex plane $\mathbb{C}$~\cite{Vekua,Ahlfors}. In
our case we have a mapping $S\left(\lambda,\bar{\lambda}\right)$
which is conformal outside the ring $G$ and quasi-conformal inside
$G$. Such mapping referred as the conformal mapping with
quasi-conformal extension. So, the quasi-conformal mappings of
this type which obey, in addition, the
properties~(\ref{B_symmetry}),~(\ref{S_0})
and~(\ref{parity_circle}) provide us with the solutions of the
plane eikonal equation~(\ref{eikonal2}). In particular, wavefronts
given by $S\left(\lambda,\bar{\lambda};z,\bar{z}\right) = const$
are level sets of such quasi-conformal mappings. In more details,
the interconnection between quasi-conformal mappings and
geometrical optics on the plane will be discussed elsewhere.

\subsection{The dVN hierarchy.}
\label{sec_QC_dVN} In this section we will apply the
quasiclassical $\bar{\partial}$-dressing method to the dVN
hierarchy. For this purpose we consider again the nonlinear
$\bar{\partial}$-problem~(\ref{Beltrami_nonlinear}),
(\ref{kernel}) on the ring $G$ with the
constraints~(\ref{B_symmetry}) and~(\ref{parity_circle}) and
introduce independent variables $x_{n}$, $\bar{x}_{n}$  via
\begin{equation}
\label{asymp_S0} S_{0} = \sum_{n=1}^{\infty}x_{n} \lambda^{2n-1} +
\sum_{n=1}^{\infty}\bar{x}_{n} \lambda^{-2n+1}.
\end{equation}
The derivatives $S_{x_{n}}$ and $S_{\bar{x}_{n}}$ obey the linear
Beltrami equation~(\ref{Beltrami}), and using the Vekua's theorem
one can construct an infinite set of equations of the form
\begin{equation}
\label{hierarchy} \Omega\left(x_{n},\bar{x}_{n},
S_{x_{n}},S_{\bar{x}_{n}} \right) = 0.
\end{equation}
Repeating the procedure described in the previous section, one
gets the plane eikonal equation for $z=x_{1}$. In a similar
manner, for the variable $\tau = x_{2} = \bar{x}_{2}$, taking into
account that $S_{\tau} = \lambda^{3} + \frac{1}{\lambda^{3}} +
\tilde{S}_{\tau}$, one obtains the equation
\begin{equation}
\label{cubic} S_{\tau} = S_{z}^{3}+S_{\bar{z}}^{3}+V S_{z} +
\bar{V} S_{\bar{z}}
\end{equation}
where $V = - 3 S_{1z}^{(\infty)} = - 3
\partial_{\bar{z}}^{-1}u_{z}$. Evaluating the terms of the order
$\lambda^{-1}$ in the both sides of equation~(\ref{cubic}), one
gets the dVN equation
\begin{equation}
\label{dVN} u_{\tau} = - 3 \left(u \partial_{\bar{z}}^{-1} u_{z}
\right)_{z} - 3 \left(u
\partial_{z}^{-1}u_{\bar{z}} \right)_{\bar{z}}.
\end{equation}
In similar manner, setting $\tau = x_{3} = \bar{x}_{3}$ one has
\begin{equation}
\label{quintic} \varphi =  S_{z}^{5} + S_{\bar{z}}^{5} + a_{3}
S_{z}^{3} + \bar{a}_{3} S_{\bar{z}}^{3} + a_{1} S_{z} +
\bar{a}_{1} S_{\bar{z}}
\end{equation}
and consequently the equations
\begin{subequations}
\begin{align}
&u_{\tau} = \left(a_{1} u \right)_{z} + \left(\bar{a}_{1} u
\right)_{\bar{z}} \\
&5 u_{z} + a_{3\bar{z}} = 0\\
&a_{1\bar{z}} + u a_{3z} + 3 a_{3} u_{z} = 0
\end{align}
\end{subequations}
Considering the higher variables $\tau_{n} = x_{n} = \bar{x}_{n}$,
($n = 2, 3, \dots$), one constructs the whole dVN hierarchy. It is
a straightforward to check that the
constraint~(\ref{parity_circle}) is compatible with
equations~(\ref{cubic}) and higher ones. So, the
$\bar{\partial}$-dressing scheme under this constraint provides us
with the real-valued solutions of the dVN hierarchy.

If one relaxes the reality condition for $u$ (i.e. does not impose
the constraint~(\ref{parity_circle})), then there is a wider
family of integrable deformations of the plane eikonal equation
with complex refractive index. To build these deformations one
again considers the
$\bar{\partial}$-problem~(\ref{Beltrami_nonlinear}), chooses the
domain $G$ as the ring (as before), imposes the
$B$-constraint~(\ref{B_symmetry}), but now chooses $S_{0}$ as
follows
\begin{equation}
\label{asymp_S0_general} S_{0} = \sum_{n=1}^{\infty}x_{n}
\lambda^{2n-1} + \sum_{n=1}^{\infty} y_{n} \lambda^{-2n+1}
\end{equation}
where $y_{1} = \bar{x}_{1} = \bar{z}$. Repeating the previous
construction, one gets again the eikonal
equation~(\ref{eikonal2}), but now with a complex-valued $u$.
Considering the derivatives $S_{x_{n}}$, $S_{y_{n}}$ with
$n=2,3,\dots$, one obtains the two set of equations
\begin{eqnarray}
\label{higher_x} &&S_{x_{n}}=\sum_{m=1}^{n} u_{m}(x_{n},y_{n})
\left(S_{z}
\right)^{2m-1},~~~~~~n = 2,3,\dots \\
\label{higher_y} &&S_{y_{n}}=\sum_{m=1}^{n} v_{m}(x_{n},y_{n})
\left(S_{\bar{z}} \right)^{2m-1},~~~~~~n = 2,3,\dots
\end{eqnarray}
Equations~(\ref{eikonal2}) and~(\ref{higher_x}) give rise to the
hierarchy of equations
\begin{equation}
\label{higher_x_2} u_{x_{n}} =
F_{n}\left(u,u_{z},u_{\bar{z}}\right),
\end{equation}
the simplest of which is of the form
\begin{equation}
\label{dVN_x} u_{x_{2}} = -3\left(u
\partial_{\bar{z}}^{-1}u_{z}\right)_{z}
\end{equation}
where
\begin{equation}
S_{x_{2}} = S_{z}^{3} + V S_{z}.
\end{equation}
Equations~(\ref{eikonal2}) and~(\ref{higher_y}) generate the
hierarchy of equations $u_{y_{n}} = \tilde{F}_{n}$, the simplest
of which is given by
\begin{equation}
\label{dVN_y} u_{y_{2}} = - 3 \left(u
\partial_{z}^{-1}u_{\bar{z}}\right)_{\bar{z}},
\end{equation}
and
\begin{equation}
S_{y_{2}} = S_{\bar{z}}^{3} + \bar{V} S_{\bar{z}}.
\end{equation}
For both of these hierarchies the quantity $\int \int_{\mathbb{C}}
u dz \wedge d\bar{z}$ is the integral of motion as for the dVN
hierarchy. It is easy to see that equations~(\ref{dVN_x})
and~(\ref{dVN_y}) imply the dVN equation~(\ref{dVN}) for the
variable $\tau = \left(x_{2}+y_{2}\right)/2$.

\subsection{Characterization of $\bar{\partial}$-data.}
\label{sec_characterization}

As we have seen, the constraints~(\ref{B_symmetry})
and~(\ref{parity_circle}) guarantee that one will get the eikonal
equation~(\ref{eikonal2}) with real valued $u$. In this section we
will discuss the characterization conditions for
$\bar{\partial}$-data $W\left(\lambda,\bar{\lambda},
S_{\lambda}\right)$ which provide us with such result. In general,
if one considers $\bar{\partial}$-problem with a kernel defined in
a ring and the function $S_{0}$ singular at two points, e.g.
$\lambda =0$ and $\lambda = \infty$, without
$B$-constraints~(\ref{B_symmetry}), one constructs the
dispersionless Laplace hierarchy \cite{Konopelchenko4} associated
with the quasiclassical limit of the Laplace equation, i.e. with
the equation

\begin{equation}
\label{magnetic} S_{z}(\lambda,z,\bar{z})
S_{\bar{z}}(\lambda,z,\bar{z}) - a S_{\bar{z}} = u(z,\bar{z}),~~~~
\forall \lambda \in \mathbb{C}
\end{equation}
where $a = \partial_{z}\tilde{S}(0,0)$ and $u = 1 +
\partial_{\bar{z}}S_{1}^{(\infty)}$. The eikonal
equation~(\ref{eikonal}) is obtained as a reduction of
equation~(\ref{magnetic}) taking $\tilde{S}(0,0)$ independent on
$z$. In fact, the $B$ condition~(\ref{B_symmetry}), producing
$\tilde{S}(0,0)=0$ realizes this reduction. In what follows we
will discuss how one has to choose the $\bar{\partial}$-data in a
way to construct the dVN hierarchy directly. We will find
constraints which are dispersionless analog of the constraints
found in \cite{Grinevich}, which specify two-dimensional
Schr\"odinger equation with real-valued potential. In particular,
we will see that it is possible to weaken slightly the
$B$-condition~(\ref{B_symmetry}), since the value $\tilde{S}(0,0)$
is fixed up to a constant by dVN reduction.

In the following we will focus on solutions of $\bar{\partial}$
problem of the form $S = S_{0}+\tilde{S}$, where $S_{0}$ has
polynomial singularities at $\lambda = 0$ and $\lambda = \infty$
and $\tilde{S}$ is holomorphic at these points and such that
\begin{eqnarray}
\label{infinity_limit} && \lim_{\lambda =
\infty}\tilde{S}\left(\lambda,\bar{\lambda}\right)
  = 0, \\
\label{zero_limit} && \lim_{\lambda =
0}\tilde{S}\left(\lambda,\bar{\lambda}\right)
  = \tilde{S}(0,0,z,\bar{z}) = -i v(z,\bar{z}),
\end{eqnarray}

{\lemma \label{lemma1} Let the kernel $W$ in
equation~\eqref{Beltrami_nonlinear} satisfies the assumptions of
the Vekua's theorem, and let $S$ be its solution. The condition
\begin{equation}
\label{parity_S} S\left(\lambda,\bar{\lambda}\right) =
-S\left(-\lambda,-\bar{\lambda}\right) + const
\end{equation}
is verified if and only if
\begin{equation}
\label{parity_W}
W'\left(\lambda,\bar{\lambda},S_{\lambda}\left(\lambda,\bar{\lambda}
\right) \right) =
W'\left(-\lambda,-\bar{\lambda},-S_{\lambda}\left(-\lambda,-\bar{\lambda}
\right) \right)
\end{equation}}

{\proof The condition~(\ref{parity_W}) is a necessary one. Indeed,
starting from~(\ref{Beltrami_nonlinear}), one obtains

\begin{eqnarray}
\label{first_one} \der{}{\bar{\lambda}}\left(S_{z}
\left(\lambda,\bar{\lambda} \right) \right) =
W'\left(\lambda,\bar{\lambda},
S_{\lambda}\left(\lambda,\bar{\lambda}\right)\right)
\der{}{\lambda}\left(S_{z}\left(\lambda,\bar{\lambda}\right)
\right) \\
\label{last_one} \der{}{\bar{\lambda}}\left(S_{z}
\left(-\lambda,-\bar{\lambda} \right) \right) =
W'\left(-\lambda,-\bar{\lambda},
-S_{\lambda}\left(-\lambda,-\bar{\lambda} \right)\right)
\der{}{\lambda}\left(S_{z}\left(-\lambda,-\bar{\lambda}\right)
\right)
\end{eqnarray}
Exploiting the condition~(\ref{parity_S}) in~(\ref{last_one}), one
gets the equality~(\ref{parity_W}).

The condition~(\ref{parity_W}) is sufficient. Indeed, let us
introduce the function
\begin{equation}
\label{Phi} \Phi\left(\lambda,\bar{\lambda},z,\bar{z}\right) =
S_{z}\left(\lambda,\bar{\lambda}\right) +
S_{z}\left(-\lambda,-\bar{\lambda} \right).
\end{equation}
Using the equations~(\ref{infinity_limit}) and~(\ref{zero_limit}),
one has
\begin{eqnarray}
&&\lim_{\lambda \rightarrow
    \infty}\Phi\left(\lambda,\bar{\lambda},z,\bar{z} \right) =
    0, \\
&&\lim_{\lambda \rightarrow
    0}\Phi\left(\lambda,\bar{\lambda},z,\bar{z} \right) = -2 i
    v_{z}\left(z,\bar{z} \right).
\end{eqnarray}
Both terms in the right hand side of~(\ref{Phi}) satisfy the
Beltrami equation~(\ref{first_one}) and~(\ref{last_one})
respectively, from which, exploiting the
constraint~(\ref{parity_W}), one concludes that
\begin{equation}
\der{\Phi}{\bar{\lambda}}\left(\lambda,\bar{\lambda},z,\bar{z}
\right) =
W'\left(\lambda,\bar{\lambda},S_{\lambda}\left(\lambda,\bar{\lambda}
\right) \right)
\der{\Phi}{\lambda}\left(\lambda,\bar{\lambda},z,\bar{z} \right).
\end{equation}
The function $\Phi$ is a solution of the Beltrami equation
vanishing at $\lambda \rightarrow \infty$. So, $\Phi$ vanishes
identically on whole $\lambda$-plane, so
\begin{equation}
\label{identity1} \der{S}{z}\left(\lambda,\bar{\lambda}\right) = -
\der{S}{z}\left(-\lambda,-\bar{\lambda} \right).
\end{equation}
In particular $\Phi\left(0,0,z,\bar{z} \right)= -2 i v_{z}\left(
z,\bar{z}\right) \equiv 0$ ,that is $v_{z}\left(z,\bar{z}\right)
\equiv 0$. Analogously it is possible to demonstrate that
\begin{equation}
\label{identity2}
\der{S}{\bar{z}}\left(\lambda,\bar{\lambda}\right) = -
\der{S}{\bar{z}}\left(-\lambda,-\bar{\lambda} \right).
\end{equation}
Hence $v_{\bar{z}}\left( z,\bar{z}\right) \equiv 0$.
Equations~(\ref{identity1}) and~(\ref{identity2}) lead to the
relation~(\ref{parity_S}) $\Box$
%end proof

{\lemma \label{lemma2} Let $S\left(\lambda,\bar{\lambda},z,\bar{z}
\right)$ be a solution of $\bar{\partial}$
problem~\eqref{Beltrami_nonlinear} such
that~\eqref{infinity_limit} and~\eqref{zero_limit} are verified.
Then the condition
\begin{equation}
\label{unit_inversion} \bar{S}\left(\lambda,\bar{\lambda}\right) =
  S\left(\frac{1}{\bar{\lambda}}, \frac{1}{\lambda} \right)+iv(z,\bar{z})
\end{equation}
is verified if and only if
\begin{equation}
\label{kernel_inversion}
 \lambda^{2}
  \overline{W}\left(\lambda,\bar{\lambda},
  S_{\lambda}\left(\lambda,\bar{\lambda} \right) \right) = -
  W\left(\frac{1}{\bar{\lambda}},\frac{1}{\lambda}, -
  \bar{\lambda}^{2}
  S_{\bar{\lambda}}\left(\frac{1}{\bar{\lambda}},\frac{1}{\lambda}
  \right) \right).
\end{equation}}

{\proof The condition~(\ref{kernel_inversion}) is a necessary one.
Let
  us consider the complex conjugation of equation~(\ref{Beltrami_nonlinear})
\begin{equation}
\label{DBAR_conjugate}
\partial_{\lambda}\left(\bar{S}\left(\lambda,\bar{\lambda} \right)\right) =
\overline{W}\left(\lambda,\bar{\lambda},
S_{\lambda}\left(\lambda,\bar{\lambda} \right) \right).
\end{equation}
Since
\begin{equation}
\bar{S}\left(\lambda,\bar{\lambda} \right) =
S\left(\frac{1}{\bar{\lambda}},\frac{1}{\lambda} \right) + i
v(z,\bar{z}) = S\left(\xi,\bar{\xi} \right) + i v(z,\bar{z})
\end{equation}
where $\xi = \bar{\lambda}^{-1}$ and $\bar{\xi} = \lambda^{-1}$,
the left hand side of~(\ref{DBAR_conjugate}) can be written as
follows
\begin{eqnarray}
\partial_{\lambda}\left(S\left(\xi,\bar{\xi} \right) \right) &=&
\der{\bar{\xi}}{\lambda} \der{S}{\bar{\xi}}\left(\xi,\bar{\xi}
\right)=-\frac{1}{\lambda^{2}}
W\left(\xi,\bar{\xi},\partial_{\xi}S\left(\xi,\bar{\xi} \right)
\right) = \nonumber \\
&=& - \frac{1}{\lambda^{2}} W\left(\frac{1}{\bar{\lambda}},
\frac{1}{\lambda}, -\bar{\lambda}^{2}
\partial_{\bar{\lambda}}S\left(\frac{1}{\bar{\lambda}},\frac{1}{\lambda}
\right) \right),
\end{eqnarray}
that provides us with equation~(\ref{unit_inversion}).

The condition~(\ref{kernel_inversion}) is a sufficient one. The
$\bar{\partial}$-equation~(\ref{Beltrami_nonlinear}) written in
terms of the variables $\xi = \bar{\lambda}^{-1}$ and $\bar{\xi} =
\lambda^{-1}$
\begin{equation}
S_{\bar{\xi}} = W\left(\xi,\bar{\xi},S_{\xi}\left(\xi,\bar{\xi}
\right) \right)
\end{equation}
is equivalent to
\begin{equation}
\label{DBAR_equiv} \lambda^{2}
S_{\lambda}\left(\frac{1}{\bar{\lambda}},\frac{1}{\lambda} \right)
= -
W\left(\frac{1}{\bar{\lambda}},\frac{1}{\lambda},-\bar{\lambda}^{2}S_{\bar{\lambda}}\left(\frac{1}{\bar{\lambda}},\frac{1}{\lambda}
\right) \right).
\end{equation}
Using equation~(\ref{DBAR_conjugate}), multiplied by
$\lambda^{2}$, and equation~(\ref{kernel_inversion}), one
concludes that
\begin{equation}
\label{parity_circle_der}
\partial_{\lambda}\bar{S}\left(\lambda,\bar{\lambda}\right) =
\partial_{\lambda} S\left(\frac{1}{\bar{\lambda}},\frac{1}{\lambda} \right).
\end{equation}
Integrating~(\ref{parity_circle_der}), one gets
\begin{equation}
\label{jump} \bar{S}\left(\lambda,\bar{\lambda} \right) =
S\left(\frac{1}{\bar{\lambda}},\frac{1}{\lambda} \right)+
\tilde{v}\left(z,\bar{z} \right).
\end{equation}

Let us note that $\tilde{v}$ cannot depend on $\bar{\lambda}$
since the function $S$ is meromorphic outside the ring. Now,
evaluating the equality~(\ref{jump}) at $\lambda=0$ and $\lambda
\rightarrow \infty$, one obtains
\begin{equation}
\tilde{v}\left(z,\bar{z} \right) =
\overline{\tilde{S}}(0,0,z,\bar{z}) = - \tilde{S}(0,0,z,\bar{z}).
\end{equation}
So, $\tilde{v}$ is a purely imaginary function
\begin{equation}
\tilde{v}\left(z,\bar{z}\right) = i v\left(z,\bar{z} \right)
\end{equation}
where $v$ is a real valued function. This complete the proof
$\Box$.

Note that $\tilde{v}_{z} = i v_{z}(z,\bar{z}) = -a$, in other
words, it is the coefficient in front of the ``magnetic'' term in
equation~(\ref{magnetic}). When it vanishes (i.e. $a=0$), one has
the pure potential equation~(\ref{magnetic}), that is the eikonal
equation.

Combining together the lemmas~(\ref{lemma1}) and~(\ref{lemma2}),
one gets the following theorem: {\bf Theorem} {\em If the
$\bar{\partial}$-data $W$ of the
  $\bar{\partial}$-equation~\eqref{Beltrami_nonlinear} obey the constraints
\begin{eqnarray}
\lambda^{2}
\overline{W}\left(\lambda,\bar{\lambda},S_{\lambda}\left(\lambda,\bar{\lambda}
\right) \right) = -
W\left(\frac{1}{\bar{\lambda}},\frac{1}{\lambda},-\bar{\lambda}^{2}S_{\bar{\lambda}}\left(\frac{1}{\bar{\lambda}},\frac{1}{\lambda}\right)
\right), \\
W'\left(\lambda,\bar{\lambda},S_{\lambda}\left(\lambda,\bar{\lambda}\right)
\right) =
W'\left(-\lambda,-\bar{\lambda},-S_{\lambda}\left(-\lambda,-\bar{\lambda}\right)\right),
\end{eqnarray}
then this $\bar{\partial}$-problem provides us with the eikonal
equation with real-valued refractive index.
%end theorem
}

\section{Concluding remarks.}
\label{sec_conclusions} Most results in nonlocal nonlinear optics,
except very few interesting cases~\cite{Snyder,KrolikExact}, have
been obtained previously by numerical analysis of the nonlocal NLS
equation (see e.g.~\cite{Krolik1,Krolik2,Krolik3}). We have
demonstrated the existence of integrable regimes for the Cole-Cole
media in the geometric optics limit. In this case many
calculations can be performed analytically. Of course, this type
of model provides us with certain approximation for the phase of
the electric field and of intensity. As it happens for the
helicoidal wavefronts, the model breaks along the lines where the
phase is singular. Then, even if our approach allows us to analyze
exactly the properties of wavefronts and the structure of their
dislocations, the approximation for intensity is no more accurate.
The helicoid is a particular meaningful example of this fact.
Indeed, our geometric optics model gives a blows up instead of a
vanishing intensity~\cite{MoroSpie}. As observed above, the
intensity of light beam vanishes due to the no more negligible
wave corrections along the axis of helicoid, where the phase is
singular. These vortex type structures are of interest from the
phenomenological point of view because the are often associated
with dark solitons in nonlocal medi~\cite{KrolikDark}a. So, one
may expect that the singularities of the phase may give a clue of
possible existence of particular interesting structures just like
dark solitons.

We also have demonstrated that in the Beltrami sector there exist
self-guided light beams exhibiting nontrivial singular phase
structures which were not considered before. An accurate
description of these wavefront dislocations could gives useful
indications about the physical properties of light beams beyond
our model.
\\

\noindent {\bf Acknowledgments.} Supported in part by COFIN PRIN
``Sintesi" 2004.

\end{document}